\documentclass[12pt]{article}
\usepackage[latin9]{inputenc}
\usepackage{amsmath}
\usepackage{amssymb}
\usepackage[unicode=true,pdfusetitle,
 bookmarks=true,bookmarksnumbered=false,bookmarksopen=false,
 breaklinks=false,pdfborder={0 0 0},pdfborderstyle={},backref=false,colorlinks=false]
 {hyperref}

\makeatletter

\InputIfFileExists{t2aenc.def}{}{%
  \errmessage{File `t2aenc.def' not found: Cyrillic script not supported}}
\DeclareRobustCommand{\cyrtext}{%
  \fontencoding{T2A}\selectfont\def\encodingdefault{T2A}}
\DeclareRobustCommand{\textcyr}[1]{\leavevmode{\cyrtext #1}}

\DeclareTextSymbolDefault{\textquotedbl}{T1}

\numberwithin{equation}{section}

\usepackage{amsfonts}
\usepackage{ytableau} 
\usepackage[square,numbers,sort&compress]{natbib}

\makeatother

\begin{document}
\begin{flushright}
FIAN/TD/2020-19\\
 
\par\end{flushright}

\vspace{0.5cm}
 
\begin{center}
\textbf{\large{}Off-shell higher-spin fields in $AdS_{4}$ and external
currents}{\large\par}
\par\end{center}

\begin{center}
\vspace{0.1cm}
\textbf{N.G.~Misuna}\\
 \vspace{0.5cm}
\textit{Tamm Department of Theoretical Physics, Lebedev Physical Institute,}\\
 \textit{ Leninsky prospect 53, 119991, Moscow, Russia}\\
\par\end{center}

\begin{center}
\vspace{0.6cm}
misuna@lpi.ru \\
\par\end{center}

\vspace{0.4cm}

\begin{abstract}
\noindent We construct an unfolded system for off-shell fields of
arbitrary integer spin in $4d$ anti-de Sitter space. To this end
we couple an on-shell system, encoding Fronsdal equations, to external
Fronsdal currents for which we find an unfolded formulation. We present
a reduction of the Fronsdal current system which brings it to the
unfolded Fierz-Pauli system describing massive fields of arbitrary
integer spin. Reformulating off-shell higher-spin system as the set
of Schwinger\textendash Dyson equations we compute propagators of
higher-spin fields in the de Donder gauge directly from the unfolded
equations. We discover operators that significantly simplify this
computation, allowing a straightforward extraction of wave equations
from an unfolded system.

\newpage{}

\tableofcontents{}
\end{abstract}

\section{Introduction}

Higher-spin (HS) gravity represents a theory of interacting massless
fields of all spins. Up to date the most complete formulation of higher-spin
gravity is provided by Vasiliev theory \cite{vas1,vas2}. It represents
a set of first-order generating equations that encode classical equations
of motion of HS fields. This method of formulating a theory in the
form of first-order constraints on exterior forms is referred to as
an unfolded dynamics approach \cite{unf2}. It provides a coordinate-independent
and manifestly gauge-invariant description of the model under consideration.
Both these features are substantial when dealing with HS theory, because
it includes gravity (a massless spin-2 field) and possesses an infinite-dimensional
HS gauge symmetry. But extracting physical quantities from Vasiliev
equations represents a complicated (both technically and conceptually)
problem. By now only cubic HS vertices have been completely derived
and analyzed \cite{cub1,key-4,key-5,key-6,key-7,cub2}, and also some
partial results on quartic and quintic vertices are available \cite{quart1,key-10,quart2}.

One of the central difficulties of HS gravity is that its full nonlinear
action is unknown. This prevents one from straightforward verification
of various $AdS/CFT$ conjectures relating HS gravity to boundary
conformal vectorial models \cite{AdSCFT1,key-13,key-14,key-15,key-16,AdSCFT2}
and from systematic study of quantum features of HS gravity. A possible
way to the action was proposed in \cite{ActionsCharges}, where it
was shown that all nontrivial gauge-invariant functionals of the unfolded
system are in one-to-one correspondence with cohomologies of a certain
operator $Q$ determined by unfolded equations. An unfolded system
is said to lie off-shell if all unfolded equations only express descendant
fields in terms of primaries, while primary fields remain unconstrained.
Then $Q$-cohomologies of the off-shell unfolded system list all candidates
for gauge-invariant actions of the theory in question. However, a
direct calculation of $Q$-cohomologies of Vasiliev equations (and
their putative off-shell completion) seems to be unfeasible. Even
for the simplest theories $Q$-cohomology problem requires some efforts
(see e.g. \cite{WZ} where $Q$-cohomology analysis was carried out
for Wess-Zumino model).

Other alternatives for an action principle of HS gravity were also
put forward, see e.g. \cite{act1,act2,act3,act4,act5,act6,act7,act8}.
And in spite of the lack of the full canonical action, certain quantum
calculations in HS gravity were successfully performed. These include
e.g. evaluations of a 1-loop vacuum partition function of HS gravity
in different geometries and comparing results with $AdS/CFT$ predictions
\cite{Z1,Z2,Z3,Z4,Z5,Z6,Z7,Z8,Z9,Z10}, computations of some $AdS$-amplitudes
(including loop corrections) for HS fields \cite{amp0,amp1,amp2,amp3,amp4,amp5,amp6,amp7},
study of the chiral HS gravity which revealed various cancellations
of UV divergences and one-loop finiteness \cite{chir1,chir2,chir3,chir4,chir5,chir6}
etc.

In \cite{misuna} a different approach to the problem of action and
quantization of HS gravity was proposed. It was argued that the procedure
of the off-shell completion of a given unfolded system amounts to
switching on external sources for all primary fields. Then the resulting
off-shell system can be reformulated as the set of Schwinger\textendash Dyson
functional equations of the quantized theory. This allows one to directly
proceed to the systematic computation of the quantum generating functional
and correlation functions. In \cite{misuna} this was illustrated
by constructing an off-shell system for free HS fields in $4d$ Minkowski
space.

This paper represents the first step in the program of quantization
of Vasiliev theory within the unfolded framework. Here we present
an off-shell unfolded system for free massless bosonic HS fields in
$4d$ anti-de Sitter space. We start with the unfolded on-shell HS
equations, which represent a linear limit of the full Vasiliev equations,
and consistently couple them to the external Fronsdal HS currents.
For these currents we construct an unfolded system. It turns out that
this HS current system allows a simple reduction to the unfolded system
for Fierz-Pauli massive HS fields. Resulting description of the massive
HS fields is non-gauge, providing an alternative to unfolding massive
HS fields in \cite{mass1,key-21,key-23,key-24,key-25,key-26,key-27,mass2}
by composing them from the couple of massless ones with all necessary
helicities. Using an off-shell completed HS system, we compute propagators
for massless HS fields in the de Donder gauge. The procedure of computation
gets substantially simplified due to the application of \textquotedbl conjugate
operators\textquotedbl{} $D^{*}$ we found, which allows one to directly
extract wave equations for component fields from the unfolded system.

The paper is organized as follows. In Section \ref{SEC_UDA} we introduce
and discuss necessary concepts and features of the unfolded dynamics
approach and explain our method of constructing off-shell completion.
In Section \ref{SEC_HS_UNFOLDED} we consider an on-shell unfolded
system for free HS fields in $AdS_{4}$, the so-called Central On-Mass-Shell
Theorem, which is a starting point for our analysis, and present a
quick way to recover Fronsdal equations from it. Section \ref{SEC_SPIN_0}
is devoted to the detailed analysis of an off-shell completion of
a scalar field, including consideration of arbitrary mass-shell reductions
and calculation of a bulk-to-bulk propagator, in order to illustrate
our idea. In Section \ref{SEC_FRNSDL_CRRNT} we construct an unfolded
system describing Fronsdal HS currents and discuss its on-shell reduction,
leading to the system for the Fierz-Pauli HS fields of arbitrary mass.
Finally, in Section \ref{SEC_OFF_SHELL_SPIN_S} we couple unfolded
Fronsdal currents to the Central On-Mass-Shell Theorem, thus formulating
an off-shell unfolded system for HS fields in $AdS_{4}$, and make
use of it in order to calculate bulk-to-bulk propagators of Fronsdal
HS fields in the de Donder gauge. Section \ref{SEC_CONCLUSIONS} contains
our conclusions. In Appendix A notations and conventions used throughout
the paper are collected. In the second Appendix B we present a different
off-shell extension for HS fields, which contains simpler equations
for HS currents but leads to non-diagonal coupling of these currents
to Fronsdal fields.

\section{Essentials of the unfolded dynamics approach\label{SEC_UDA}}

Unfolded formulation \cite{unf1,vas1,vas2,unf2,ActionsCharges} of
the theory implies its representation as the set of equations of the
form
\begin{equation}
\mathrm{d}W^{A}(x)+G^{A}(W)=0,\label{unf_eq}
\end{equation}
where $\mathrm{d}$ is the de Rham differential on the spacetime manifold
$M^{d}$ with local coordinates $x$; $W^{A}(x)$ are unfolded fields
representing spacetime exterior forms, with $A$ denoting all their
indices; $G^{A}(W)$ is built from exterior products of $W$ (we will
omit the wedge symbol throughout the paper). The nilpotency of the
de Rham differential $\mathrm{d}^{2}\equiv0$ imposes a consistency
condition on $G$
\begin{equation}
G^{B}\dfrac{\delta G^{A}}{\delta W^{B}}\equiv0,\label{unf_consist}
\end{equation}
which plays the crucial role in the unfolded analysis. An unfolded
system \eqref{unf_eq} is manifestly invariant under a set of infinitesimal
gauge transformations
\begin{equation}
\delta W^{A}=\mathrm{d}\varepsilon^{A}(x)-\varepsilon^{B}\dfrac{\delta G^{A}}{\delta W^{B}}\label{unf_gauge_transf}
\end{equation}
(in checking the invariance one should make use of \eqref{unf_consist}).
A gauge parameter $\varepsilon^{A}(x)$, representing a rank-$(n-1)$
form, is associated with a gauge transformation generated by rank-$n$
unfolded field $W^{A}$. 0-forms do not give rise to gauge symmetries
and are transformed only by gauge transformations of higher-rank fields
due to the second term in \eqref{unf_gauge_transf}.

The spacetime geometry in the unfolded approach is described by a
generalized 1-form connection $\Omega=\mathrm{d}x^{\underline{a}}\Omega_{\underline{a}}^{A}(x)T_{A}$
that takes values in the Lie algebra of spacetime symmetries with
generators $T_{A}$. Maximally symmetric gravitational background
arises via imposing zero-curvature condition on $\Omega$
\begin{equation}
\mathrm{d}\Omega+\frac{1}{2}[\Omega,\Omega]=0\label{flat_conn}
\end{equation}
 (square brackets stand for the Lie-algebra commutator). Fixing some
particular solution $\Omega_{0}$ to this equation breaks a gauge
symmetry
\begin{equation}
\delta\Omega=\mathrm{d}\varepsilon(x)+[\Omega,\varepsilon]
\end{equation}
 to the leftover global symmetry $\varepsilon_{glob}$ that leaves
$\Omega_{0}$ invariant and thus satisfies
\begin{equation}
\mathrm{d}\varepsilon_{glob}+[\Omega_{0},\varepsilon_{glob}]=0.\label{glob_symm}
\end{equation}

In this paper we deal with $d=4$ anti-de Sitter space $AdS_{4}$,
so we introduce a corresponding connection of its symmetry algebra
$so(3,2)$
\begin{equation}
\Omega^{AdS}=e^{\alpha\dot{\beta}}P_{\alpha\dot{\beta}}+\Omega^{\alpha\beta}M_{\alpha\beta}+\bar{\Omega}^{\dot{\alpha}\dot{\beta}}\bar{M}_{\dot{\alpha}\dot{\beta}},\label{AdS_connection}
\end{equation}
where $P_{\alpha\dot{\alpha}}$, $M_{\alpha\beta}$ and $\bar{M}_{\dot{\alpha}\dot{\beta}}$
represent generators of spacetime translations and (selfdual and anti-selfdual
part of) rotations, $e^{\alpha\dot{\beta}}$ and $\Omega^{\alpha\beta}$
($\bar{\Omega}^{\dot{\alpha}\dot{\beta}}$) are 1-forms of vierbein
and Lorentz connection. Expansion of \eqref{flat_conn} in generators
gives
\begin{align}
 & \mathrm{d}e^{\alpha\dot{\beta}}+\Omega^{\alpha}\text{}_{\gamma}e^{\gamma\dot{\beta}}+\bar{\Omega}^{\dot{\beta}}\text{}_{\dot{\gamma}}e^{\alpha\dot{\gamma}}=0,\label{ads1}\\
 & \mathrm{d}\Omega^{\alpha\beta}+\Omega^{\alpha}\text{}_{\gamma}\Omega^{\gamma\beta}=-\lambda^{2}E^{\alpha\beta},\\
 & \mathrm{d}\bar{\Omega}^{\dot{\alpha}\dot{\beta}}+\bar{\Omega}^{\dot{\alpha}}\text{}_{\dot{\gamma}}\bar{\Omega}^{\dot{\gamma}\dot{\beta}}=-\lambda^{2}\bar{E}^{\dot{\alpha}\dot{\beta}}.\label{ads2}
\end{align}
Here $E^{\alpha\beta}=e^{\alpha}{}_{\dot{\gamma}}e^{\beta\dot{\gamma}}$,
$\bar{E}^{\dot{\alpha}\dot{\beta}}=e_{\gamma}\text{}^{\dot{\alpha}}e^{\gamma\dot{\beta}}$
are basis 2-forms and $\lambda$ denotes an inverse $AdS$-radius.
If now one chooses some particular solution to \eqref{ads1}-\eqref{ads2},
i.e. fixes some coordinate frame in $AdS_{4}$, then solutions of
\eqref{glob_symm} will provide an explicit realization of $AdS$
global symmetries in these coordinates.

To illustrate the unfolded approach in more detail and explain our
notations let us consider an unfolded formulation of a free massless
scalar field. The scalar field is described by an unfolded module
\begin{equation}
C(Y|x)=\sum_{N=0}^{\infty}C_{N}(Y|x)=\sum_{N=0}^{\infty}C_{\alpha(N),\dot{\alpha}(N)}(x)y^{\alpha_{1}}...y^{\alpha_{N}}\bar{y}^{\dot{\alpha}_{1}}...\bar{y}^{\dot{\alpha}_{N}}\label{C_scalar}
\end{equation}
where $Y=(y^{\alpha},\bar{y}^{\dot{\alpha}})$ is a pair of auxiliary
commuting $sp(2)$-spinors (see Appendix A for notation details).
Unfolded equations for $C$ are
\begin{equation}
DC+ie^{\alpha\dot{\beta}}\partial_{\alpha}\bar{\partial}_{\dot{\beta}}C-ie^{\alpha\dot{\beta}}y_{\alpha}\bar{y}_{\dot{\beta}}C=0,\label{unf_scal}
\end{equation}
where
\begin{equation}
D=\frac{1}{\lambda}(\mathrm{d}+\Omega^{\alpha\beta}y_{\alpha}\partial_{\beta}+\bar{\Omega}^{\dot{\alpha}\dot{\beta}}\bar{y}_{\dot{\alpha}}\bar{\partial}_{\dot{\beta}})\label{D_definition}
\end{equation}
is a dimensionless 1-form of a Lorentz-covariant derivative. To study
the content of \eqref{unf_scal} one expands $D$ in vierbein basis
as $D=e^{\alpha\dot{\beta}}D_{\alpha\dot{\beta}}$ which yields
\begin{equation}
D_{\alpha\dot{\beta}}C+i\partial_{\alpha}\bar{\partial}_{\dot{\beta}}C-iy_{\alpha}\bar{y}_{\dot{\beta}}C=0.
\end{equation}
Contracting this with $iy^{\alpha}\bar{y}^{\dot{\beta}}$ leads to
\begin{equation}
C_{N}(Y|x)=\frac{1}{(N!)^{2}}(iy^{\alpha}\bar{y}^{\dot{\beta}}D_{\alpha\dot{\beta}})^{N}C(0|x),\label{scal_desc}
\end{equation}
while contracting with $-\frac{\lambda^{2}}{2}(D^{\alpha\dot{\beta}}-i\partial^{\alpha}\bar{\partial}^{\dot{\beta}}+iy^{\alpha}\bar{y}^{\dot{\beta}})$
produces
\begin{equation}
\square C_{N}+\lambda^{2}\left(N^{2}+2N+2\right)C_{N}=0,
\end{equation}
where $\square=-\frac{\lambda^{2}}{2}D^{\alpha\dot{\beta}}D_{\alpha\dot{\beta}}$
is the wave operator in $AdS_{4}$. So for the primary field $\phi(x)=C_{0}(0|x)$
this gives
\begin{equation}
\square\phi+2\lambda^{2}\phi=0.\label{scal_eom}
\end{equation}
Thus unfolded system \eqref{unf_scal} indeed describes a massless
scalar field $\phi(x)$ in $AdS_{4}$\footnote{As usual, in $AdS$ by massless we mean a scalar field which is conformally
coupled to the $AdS$-curvature, that fixes the mass-like term in
\eqref{scal_eom}}.

One sees from \eqref{scal_desc} that $C_{N}(Y|x)$ with $N>0$ are
descendant fields that form a tower of totally symmetrized traceless
derivatives of the primary scalar $\phi(x)$. The system is on-shell
because the primary field is subjected to the differential constraint
\eqref{scal_eom}. Global transformations \eqref{glob_symm} for $AdS$-connection
\eqref{AdS_connection} induce a massless scalar representation of
$AdS_{4}$-algebra $so(3,2)$ on the unfolded module \eqref{C_scalar}
via the general formula \eqref{unf_gauge_transf}.

The system \eqref{unf_scal} contains an infinite number of unfolded
equations on infinite number of unfolded fields. This is a typical
situation: while unfolded, a system with an infinite number of d.o.f.
(like a relativistic field) generates an infinite number of descendants
which parameterize all these d.o.f (compare with three unfolded equations
\eqref{ads1}-\eqref{ads2} determining non-dynamical $AdS$ background).
If the initial system is on-shell, the basis of unfolded descendants
is \textquotedbl incomplete\textquotedbl : some subset of possible
descendants is absent - namely those that correspond to the l.h.s.
of e.o.m. and all their differential consequences. In the scalar field
example the basis of descendants \eqref{scal_desc} contains only
traceless symmetrized derivatives, and there are no descendants containing
d'Alembertians of $\phi$, because those are fixed by the mass-shell
constraint \eqref{scal_eom}.\footnote{Also there are no descendants containing antisymmetrization of derivatives,
but this is because due to $AdS$ commutator $\left[D,D\right]\sim\Lambda$
such combinations reduce to the lower descendants from \eqref{scal_desc}.}

Thus, the problem of constructing an off-shell extension for a given
on-shell unfolded system amounts to the completion of the basis of
unfolded descendants. This can be performed at the level of the unfolded
equations by restoring all absent descendants and introducing them
to the $G^{A}$ of the on-shell system \eqref{unf_eq} in the way
compatible with the consistency requirement \eqref{unf_consist}.
However, as was proposed in \cite{misuna}, another way can be followed,
if the spectrum of the primary fields and their e.o.m. are known.
This input either can be known from the very beginning or can be gained
by carrying out the $\sigma_{-}$-analysis \cite{sigma_cohomol,ActionsCharges}
that allows for the systematic extraction of the dynamical content
of the unfolded system.

To get the idea of \cite{misuna}, consider once again a free massless
scalar field. Conventional on-shell formulation is provided by equation
\eqref{scal_eom} and unfolding this results in \eqref{unf_scal}.
Now let us couple $\phi$ to an external source $J(x)$. Then \eqref{scal_eom}
turns to
\begin{equation}
\square\phi+2\lambda^{2}\phi=J.\label{scal_eom_J}
\end{equation}
If $J$ takes some prescribed value, then \eqref{scal_eom_J} determines
a corresponding backreaction of $\phi$ to this $J$. But if $J$
is an a priori unknown function, with \eqref{scal_eom_J} being the
only relation involving it, one can treat \eqref{scal_eom_J} as the
definition of $J$. In this case the theory in question can be considered
as lying off-shell: it describes two scalar fields, $\phi(x)$ and
$J(x)$, with primary $\phi$ totally unconstrained and descendant
$J$ defined by \eqref{scal_eom_J}. Unfolding this system is equivalent
to unfolding off-shell scalar $\phi$. So to construct an off-shell
completion of the given on-shell theory one should couple it to the
non-fixed external sources, which from the standpoint of the unfolded
approach will play a role of the \textquotedbl prodigal sons\textquotedbl ,
i.e. previously absent descendants.

This method of constructing an off-shell completion has as important
advantage as it paves the way to the quantization of the theory directly
within the unfolded framework. In the conventional QFT, if the classical
e.o.m. for fields $\left\{ \varphi_{k}(x)\right\} $, arising from
the action $S$, are
\begin{equation}
\dfrac{\delta S}{\delta\varphi_{n}}(\varphi_{k})=0,\label{class_EoM}
\end{equation}
then Schwinger\textendash Dyson equations for the partition function
$Z[J]=\int\mathcal{D}\varphi\exp\left\{ iS[\varphi]-iJ_{k}\varphi^{k}\right\} $
of the quantum theory are
\begin{equation}
\dfrac{\delta S}{\delta\varphi_{n}}(i\dfrac{\delta}{\delta J_{k}})Z=J_{n}Z.\label{SchDys_eq}
\end{equation}
(Formulation of Schwinger\textendash Dyson equations for non-Lagrangian
theories, involving the construction of the so-called Lagrange anchor,
is considered in \cite{Lyakh1,Lyakh2,Lyakh3}.) So having at one's
disposal an off-shell system or equivalently an on-shell system coupled
to the external currents
\begin{equation}
\dfrac{\delta S}{\delta\varphi_{n}}(\varphi_{k})=J_{n},\label{class_EoM_J}
\end{equation}
by a substitution $\varphi_{k}\rightarrow i\frac{\delta}{\delta J_{k}}$
one arrives at Schwinger\textendash Dyson equations \eqref{SchDys_eq}
that determine the partition function $Z$ and, hence, the whole quantum
theory.

\section{Unfolded on-shell HS fields and Fronsdal equations\label{SEC_HS_UNFOLDED}}

Our aim is to provide an unfolded off-shell formulation for bosonic
HS fields in $AdS_{4}$. In conventional Lagrangian language, a massless
integer spin-$s$ field propagating in $AdS_{4}$ is described, as
was found by Fronsdal \cite{Fronsdal}, by a double-traceless rank-$s$
Lorentz tensor $\varphi_{\underline{a}(s)}(x)$ with classical e.o.m.
\begin{equation}
\square\varphi_{\underline{a}(s)}-sD_{\underline{a}}D^{\underline{b}}\varphi_{\underline{ba}(s-1)}+\frac{s(s-1)}{2}D_{\underline{a}}D_{\underline{a}}\varphi^{\underline{b}}\mbox{}_{\underline{ba}(s-2)}-\lambda^{2}(s^{2}-2s-2)\varphi_{\underline{a}(s)}-\lambda^{2}s(s-1)g_{\underline{aa}}\varphi^{\underline{b}}\mbox{}_{\underline{ba}(s-2)}=0,\label{Fronsdal_eq}
\end{equation}
and a gauge transformation law (for $s>0$)
\begin{equation}
\delta\varphi_{\underline{a}(s)}=D_{\underline{a}}\epsilon_{\underline{a}(s-1)}.
\end{equation}

In the unfolded language, the same theory is described by a 1-form
$\omega(Y|x)$ (encoding Fronsdal gauge spin-$s$ field and its first
$(s-1)$ derivatives) and 0-form $C(Y|x)$ (encoding gauge-invariant
HS curvatures and infinite towers of their descendants)
\begin{equation}
\omega(Y|x)=\sum_{n,m}\omega_{\underline{a}|\alpha(n),\dot{\beta}(m)}y^{\alpha_{1}}...y^{\alpha_{n}}\bar{y}^{\dot{\beta}_{1}}...\bar{y}^{\dot{\beta}_{m}}\mathrm{d}x^{\underline{a}},\quad C(Y|x)=\sum_{n,m}C_{\alpha(n),\dot{\beta}(m)}y^{\alpha_{1}}...y^{\alpha_{n}}\bar{y}^{\dot{\beta}_{1}}...\bar{y}^{\dot{\beta}_{m}}.\label{w_C_def}
\end{equation}
Powers of $y$ and $\bar{y}$ in \eqref{w_C_def} for a concrete spin
$s$ are not independent. To fix them we introduce following important
operators
\begin{equation}
N=y^{\alpha}\partial_{\alpha},\quad\bar{N}=\bar{y}^{\dot{\alpha}}\bar{\partial}_{\dot{\alpha}},\label{N_Nbar_definition}
\end{equation}
\begin{equation}
\varsigma=\frac{1}{2}(N+\bar{N}),\quad\tau=\frac{1}{2}(N-\bar{N}),\label{s_t_definition}
\end{equation}
so that $N$ and $\bar{N}$ count the number of $y$ and $\bar{y},$
respectively, while $\varsigma$ and $\tau$ count their half-sum
and half-difference. Then for a spin-$s$ field one has
\begin{equation}
\varsigma\omega=(s-1)\omega,\quad|\tau|C=sC.\label{w_C_restrictions}
\end{equation}
Relations \eqref{w_C_restrictions} have a simple interpretation in
terms of the Young diagrams of $AdS$-tensors that $\omega$ and $C$
span: they mean that $\omega$ includes all one- and two-row Young
diagrams with the upper row length equal to $(s-1)$, while $C$ includes
all two-row diagrams with the lower row length equal to $s$. A primary
double-traceless Fronsdal spin-$s$ field is identified with
\begin{equation}
\varphi_{\underline{a}(s)}=\omega_{\underline{a}|\alpha(s-1),\dot{\alpha}(s-1)}(e_{\underline{a}}^{\alpha\dot{\alpha}})^{s-1}.
\end{equation}
The non-gauge scalar field is degenerate from this point of view:
it does not have $\omega$, while its $C$ spans all one-row diagrams
as follows from \eqref{C_scalar}.

In terms of $\omega$ and $C$, the Fronsdal theory is described by
unfolded equations referred to as Central On-Mass-Shell Theorem \cite{unf1}
\begin{align}
 & D\omega+e^{\alpha\dot{\beta}}y_{\alpha}\bar{\partial}_{\dot{\beta}}\omega+e^{\alpha\dot{\beta}}\partial_{\alpha}\bar{y}_{\dot{\beta}}\omega=\dfrac{i}{4}\eta\bar{E}^{\dot{\alpha}\dot{\beta}}\bar{\partial}_{\dot{\alpha}}\bar{\partial}_{\dot{\beta}}C|_{N=0}+\dfrac{i}{4}\bar{\eta}E^{\alpha\beta}\partial_{\alpha}\partial_{\beta}C|_{\bar{N}=0},\label{OnShTh_1}\\
 & DC+ie^{\alpha\dot{\beta}}\partial_{\alpha}\bar{\partial}_{\dot{\beta}}C-ie^{\alpha\dot{\beta}}y_{\alpha}\bar{y}_{\dot{\beta}}C=0,\label{OnShTh_2}
\end{align}
where $\eta$ is an arbitrary unimodular $\eta\bar{\eta}=1$ phase
parameter accounting for parity breaking. These equations arise in
the linear limit of the full nonlinear Vasiliev equations \cite{vas2}
after solving for auxiliary generating variables.

Let us see how \eqref{OnShTh_1} encodes Fronsdal equation \eqref{Fronsdal_eq}.
To this end one expands $\omega$ in the vierbein basis as
\begin{equation}
\omega(Y|x)=e^{\alpha\dot{\alpha}}(x)\omega_{\alpha\dot{\alpha}}(Y|x)
\end{equation}
and then decompose a 0-form $\omega_{\alpha\dot{\alpha}}(Y|x)$ as
\begin{equation}
\omega_{\alpha\dot{\alpha}}(Y|x)=\partial_{\alpha}\bar{\partial}_{\dot{\alpha}}\phi(Y|x)+y_{\alpha}\bar{y}_{\dot{\alpha}}\widetilde{\phi}(Y|x)+\bar{y}_{\dot{\alpha}}\partial_{\alpha}\chi(Y|x)+y_{\alpha}\bar{\partial}_{\dot{\alpha}}\bar{\chi}(Y|x).\label{w_decompos}
\end{equation}
Because of the constraint \eqref{w_C_restrictions}, for constituent
0-forms one has
\begin{equation}
\varsigma\phi=s\phi,\quad\varsigma\widetilde{\phi}=(s-2)\widetilde{\phi},\quad\varsigma\chi=(s-1)\chi,\quad\varsigma\bar{\chi}=(s-1)\bar{\chi}.
\end{equation}
Contracting \eqref{w_decompos} with $y^{\alpha}\bar{y}^{\dot{\alpha}}$,
$\partial^{\alpha}\bar{\partial}^{\dot{\alpha}}$, $y^{\alpha}\bar{\partial}^{\dot{\alpha}}$
or $\bar{y}^{\dot{\alpha}}\partial^{\alpha}$ one finds
\begin{align}
 & \phi=\frac{1}{N\bar{N}}y^{\alpha}\bar{y}^{\dot{\alpha}}\omega_{\alpha\dot{\alpha}},\quad\widetilde{\phi}=\frac{1}{(N+2)(\bar{N}+2)}\partial^{\alpha}\bar{\partial}^{\dot{\alpha}}\omega_{\alpha\dot{\alpha}},\\
 & \chi=-\frac{1}{N(\bar{N}+2)}y^{\alpha}\bar{\partial}^{\dot{\alpha}}\omega_{\alpha\dot{\alpha}},\quad\bar{\chi}=-\frac{1}{(N+2)\bar{N}}\bar{y}^{\dot{\alpha}}\partial^{\alpha}\omega_{\alpha\dot{\alpha}}.
\end{align}
Plugging this back into \eqref{w_decompos} one obtains a following
resolution of the identity for $\omega_{\alpha\dot{\alpha}}$:
\begin{equation}
\omega_{\alpha\dot{\alpha}}=\mathbb{P}_{\alpha\dot{\alpha}}{}^{\beta\dot{\beta}}\omega_{\beta\dot{\beta}},\quad\mathbb{P}_{\alpha\dot{\alpha}}{}^{\beta\dot{\beta}}=\frac{1}{(N+1)(\bar{N}+1)}(\partial_{\alpha}\bar{\partial}_{\dot{\alpha}}y^{\beta}\bar{y}^{\dot{\beta}}+y_{\alpha}\bar{y}_{\dot{\alpha}}\partial^{\beta}\bar{\partial}^{\dot{\beta}}-y_{\alpha}\bar{\partial}_{\dot{\alpha}}\bar{y}^{\dot{\beta}}\partial^{\beta}-\bar{y}_{\dot{\alpha}}\partial_{\alpha}y^{\beta}\bar{\partial}^{\dot{\beta}}).\label{resol_of_identity}
\end{equation}
Four terms comprising $\mathbb{P}_{\alpha\dot{\alpha}}{}^{\beta\dot{\beta}}$
form a complete set of orthogonal projectors. They correspond to the
decomposition of $\omega_{\alpha\dot{\alpha}}(Y|x)$ in terms of the
Young diagrams for Lorentz tensors: an additional, as compared to
the Young diagrams of 1-form $\omega(Y|x)$, cell $\alpha\dot{\alpha}$
of $\omega_{\alpha\dot{\alpha}}$ can be added to the first row (first
term in \eqref{resol_of_identity}), subtracted from the first row
(second term), or added or subtracted from the second row (third/fourth
term depending on the sign of $\tau$).

Now one can construct first-order operators which, being formally
$(-2)$-forms, directly extract wave equations for constituent 0-forms
from unfolded equations on $\omega$. For $\phi$ and $\widetilde{\phi}$
they look as, respectively,
\begin{align}
D_{\phi}^{*}= & \frac{\lambda^{2}}{4\varsigma(\varsigma+1)}\left(y^{\alpha}\bar{y}^{\dot{\alpha}}D_{\alpha\dot{\alpha}}y^{\beta}\partial^{\gamma}+\frac{(1-\tau)}{(\varsigma+\tau)}\bar{y}^{\dot{\alpha}}\partial^{\alpha}D_{\alpha\dot{\alpha}}y^{\beta}y^{\gamma}+(\varsigma-\tau+1)(1+\tau)y^{\beta}y^{\gamma}\right)\dfrac{\partial^{2}}{\partial e^{\beta}\text{}_{\dot{\beta}}\partial e^{\gamma\dot{\beta}}}+h.c.\label{D*_phi}\\
D_{\widetilde{\phi}}^{*}= & \frac{\lambda^{2}}{4(\varsigma+1)(\varsigma+2)}\left(\partial^{\alpha}\bar{\partial}^{\dot{\alpha}}D_{\alpha\dot{\alpha}}y^{\beta}\partial^{\gamma}-\frac{(1+\tau)}{(\varsigma+\tau+2)}y^{\alpha}\bar{\partial}^{\dot{\alpha}}D_{\alpha\dot{\alpha}}\partial^{\beta}\partial^{\gamma}-(\varsigma-\tau+1)(1-\tau)\partial^{\beta}\partial^{\gamma}\right)\cdot\nonumber \\
 & \cdot\dfrac{\partial^{2}}{\partial e^{\beta}\text{}_{\dot{\beta}}\partial e^{\gamma\dot{\beta}}}+h.c.\label{D*_phi_tilde}
\end{align}
Here $h.c.$ operation amounts to the exchange of dotted and undotted
variables and hence $\varsigma\rightarrow\varsigma$, $\tau\rightarrow-\tau$
as follows from \eqref{s_t_definition}. Acting with \eqref{D*_phi}
and \eqref{D*_phi_tilde} on the first equation of the Central On-Mass-Shell
Theorem \eqref{OnShTh_1} produces respectively
\begin{align}
 & \square\phi+\frac{\lambda^{2}}{2s}(y^{\beta}\bar{y}^{\dot{\beta}}D_{\beta\dot{\beta}})(\partial^{\alpha}\bar{\partial}^{\dot{\alpha}}D_{\alpha\dot{\alpha}})\phi-\frac{\lambda^{2}}{2s}(y^{\beta}\bar{y}^{\dot{\beta}}D_{\beta\dot{\beta}})^{2}\widetilde{\phi}-\lambda^{2}(s^{2}-2s-2-\tau^{2})\phi=\nonumber \\
 & =\frac{i\bar{\eta}\lambda^{2}(s-1)}{4(s+1)}\left(s+1-\frac{(s-2)}{(2s-1)}(\bar{y}^{\dot{\beta}}\partial^{\beta}D_{\beta\dot{\beta}})\right)C|_{\tau=s}+h.c.\label{phi_eq}
\end{align}
\begin{equation}
\square\widetilde{\phi}-\frac{\lambda^{2}}{2s}(\partial^{\alpha}\bar{\partial}^{\dot{\alpha}}D_{\alpha\dot{\alpha}})(y^{\beta}\bar{y}^{\dot{\beta}}D_{\beta\dot{\beta}})\widetilde{\phi}+\frac{\lambda^{2}}{2s}(\partial^{\alpha}\bar{\partial}^{\dot{\alpha}}D_{\alpha\dot{\alpha}})^{2}\phi-\lambda^{2}(s^{2}+2s-2-\tau^{2})\widetilde{\phi}=0.\label{phi_tilde_eq}
\end{equation}
Considering $\tau=0$ sector, where $\phi_{\alpha(s),\dot{\alpha}(s)}$
and $\widetilde{\phi}_{\alpha(s-2),\dot{\alpha}(s-2)}$ are identified
with the traceless and trace parts of spin-$s$ Fronsdal field $\varphi_{\underline{a}(s)}$,
one sees that equations \eqref{phi_eq} and \eqref{phi_tilde_eq}
reproduce traceless and trace parts of the Fronsdal equation \eqref{Fronsdal_eq}.
Thus, Central On-Mass-Shell Theorem \eqref{OnShTh_1}-\eqref{OnShTh_2}
indeed provides an unfolded formulation for the Fronsdal theory.

The main goal of the paper is to construct an off-shell extension
of the system \eqref{OnShTh_1}-\eqref{OnShTh_2}. To this end, as
we discussed in the previous Section, one should consistently couple
it to external HS currents. From \eqref{Fronsdal_eq} it follows that
these HS currents represent gauge-invariant double-traceless fields
$J_{\underline{a}(s)}^{F}(x)$ subjected to the generalized conservation
law \cite{Fronsdal}
\begin{equation}
D^{\underline{b}}J_{\underline{ba}(s-1)}^{F}=\frac{(s-1)}{2}D_{\underline{a}}J^{F\underline{b}}\mbox{}_{\underline{ba}(s-2)}.\label{conserv_law}
\end{equation}
So the first task is to unfold the system of $J_{\underline{a}(s)}^{F}(x)$
with \textquotedbl on-shell constraint\textquotedbl{} \eqref{conserv_law}.
But let us start with a simpler degenerate spin-$0$ case, which is
helpful to illustrate the general technique.

\section{Off-shell completion of the scalar field\label{SEC_SPIN_0}}

An external current for a scalar field is another scalar field which
is unconstrained. So an unfolded system for a scalar source should
represent some deformation of \eqref{unf_scal}. First, we introduce
an unfolded module describing spin-$0$ source $J^{(0)}$ constrained
by condition $(\square+2\lambda^{2})J^{(0)}=0$,
\begin{equation}
J^{(0)}(Y|x)=\sum_{N=0}^{\infty}J_{N}^{(0)}(Y|x)=\sum_{N=0}^{\infty}J_{\alpha(N),\dot{\alpha}(N)}^{(0)}(x)(y^{\alpha})^{N}(\bar{y}^{\dot{\alpha}})^{N},
\end{equation}
\begin{equation}
DJ^{(0)}+ie\partial\bar{\partial}J^{(0)}-iey\bar{y}J^{(0)}=0.\label{J0_onshell}
\end{equation}
(From now on we omit contracted spinor indices as explained in Appendix A.)
We add $J^{(0)}$ to the r.h.s. of \eqref{unf_scal} with some coefficients
$k_{N}^{(0)}$ dependent on $N$
\begin{equation}
DC+ie\partial\bar{\partial}C-iey\bar{y}C=iey\bar{y}k_{N}^{(0)}J^{(0)}.\label{unf_scal-1}
\end{equation}
Values of $k_{N}^{(0)}$ are constrained by consistency condition
\eqref{unf_consist}, as we will see shortly. Now, in order to relax
the constraint $(\square+2\lambda^{2})J^{(0)}=0$ one can similarly
introduce \textquotedbl source for source\textquotedbl{} $J^{(1)}$
to the r.h.s. of \eqref{J0_onshell}. Then, to relax $(\square+2\lambda^{2})J^{(1)}=0$,
one has to introduce $J^{(2)}$ and so on. At the end of the day,
one arrives at an infinite sequence of sources $J(Y|x)=\sum_{n=0}^{\infty}J^{(n)}$
subjected to unfolded equations
\begin{equation}
DJ^{(n)}+ie\partial\bar{\partial}J^{(n)}-iey\bar{y}J^{(n)}=iey\bar{y}k_{N}^{(n+1)}J^{(n+1)}.\label{Ji_eq}
\end{equation}
Consistency condition \eqref{unf_consist} requires
\begin{equation}
(N+2)k_{N}^{(n)}=Nk_{N-1}^{(n)},
\end{equation}
which is the only constraint on coefficients $k_{N}^{(n)}$. There
is always certain (usually very large) freedom in the choice of coefficients
in unfolded equations, because new descendants can be introduced with
arbitrary multipliers, when first appearing in an unfolded system.
This freedom can be used to simplify the form of equations, but the
choice may affect the further analysis. We will face such a situation
below. Here we fix coefficients to be
\begin{equation}
k_{N}^{(n)}=-\frac{g\lambda^{-2}}{(N+1)(N+2)}
\end{equation}
with $g$ playing the role of the coupling constant. It is convenient
to organize all $J^{(n)}$ into a single master-source $J(Y|p|x)$
as a formal expansion in some auxiliary parameter $p$
\begin{equation}
J(Y|p|x)=\sum_{n=0}^{\infty}\frac{p^{n}}{n!}J^{(n)}(Y|x)=\sum_{n,N=0}^{\infty}\frac{p^{n}}{n!}J_{\alpha(N),\dot{\alpha}(N)}^{(n)}(x)(y^{\alpha})^{N}(\bar{y}^{\dot{\alpha}})^{N}.\label{scal_source_J}
\end{equation}
Then an unfolded system describing off-shell scalar field takes the
form
\begin{align}
 & DC+ie\partial\bar{\partial}C-iey\bar{y}C=-iey\bar{y}\frac{g\lambda^{-2}}{(N+1)(N+2)}J^{(0)},\label{off-shell_scalar_1}\\
 & DJ+ie\partial\bar{\partial}J-iey\bar{y}J=-iey\bar{y}\frac{g\lambda^{-2}}{(N+1)(N+2)}\frac{\partial}{\partial p}J.\label{off-shell_scalar_2}
\end{align}
Acting on \eqref{off-shell_scalar_1} with an operator
\begin{equation}
D_{C}^{*}=-\frac{\lambda^{2}}{2}(D^{\alpha\dot{\beta}}-i\partial^{\alpha}\bar{\partial}^{\dot{\beta}}+iy^{\alpha}\bar{y}^{\dot{\beta}})\frac{\partial}{\partial e^{\alpha\dot{\beta}}},\label{D*_C}
\end{equation}
which represents an analogue of \eqref{D*_phi} and \eqref{D*_phi_tilde}
for the second equation of the Central On-Mass-Shell Theorem \eqref{OnShTh_2},
extracts wave equations for components
\begin{equation}
\square C_{N}+\lambda^{2}(N^{2}+2N+2)C_{N}=\frac{ig}{2}y^{\alpha}\bar{y}^{\dot{\beta}}D_{\alpha\dot{\beta}}\frac{1}{(N+1)(N+2)}J_{N-1}^{(0)}+\frac{g(N+2)}{2(N+1)}J_{N}^{(0)}.
\end{equation}
For the primary field $\phi(x)=C_{0}(0|x)$ this gives
\begin{equation}
\square\phi+2\lambda^{2}\phi=gJ_{0}^{(0)}(x).\label{box_scal_J}
\end{equation}
So \eqref{off-shell_scalar_1}-\eqref{off-shell_scalar_2} indeed
describe a scalar field $\phi(x)$ coupled to an external current
$J_{0}^{(0)}(x)$ with coupling constant $g$ or, interpreted differently,
they describe an unfolded off-shell scalar field, with $J(Y|p|x)$
encoding descendants which contain d'Alembertians of the primary scalar
$\phi(x)$.

From the point of view of representation theory, the above construction
goes as follows. One starts with $SO(3,2)$-module $D(1,0)\oplus D(2,0)$,
corresponding to an $AdS_{4}$ massless scalar field (two submodules
here are due to two different boundary asymptotics of the $AdS$-scalar),
and \textquotedbl glues\textquotedbl{} it to an external current module.
This results in an indecomposable representation which contains the
external current as an invariant subspace. The peculiarity of the
scalar field is that its external current module itself represents
an infinite chain of the same gluings: all $J^{(n)}(Y|x)$ in the
expansion \eqref{scal_source_J} are isomorphic as $SO(3,2)$-modules
by construction, and, as seen from \eqref{off-shell_scalar_1}-\eqref{off-shell_scalar_2},
for any $n_{0}$ a subspace of unfolded fields
\begin{equation}
J^{(>n_{0})}:=\sum_{n=n_{0}+1}^{\infty}\frac{p^{n}}{n!}J^{(n)}(Y|x)\label{J_n0}
\end{equation}
corresponds to an invariant subspace in the dual Verma module. So
the off-shell scalar module we constructed represents an infinitely
indecomposable sequence of glued $D(1,0)\oplus D(2,0)$ modules, such
that each subsequent $D(1,0)\oplus D(2,0)$ is nested as an invariant
subspace.

Now, having in hand this off-shell system, one can move in two different
directions. First, one can perform various consistent reductions of
the unfolded module, getting rid of some part of descendants. This
leads to different on-shell theories with different equations of motion.
As we will see, on this way it is possible to describe an on-shell
scalar field with arbitrary mass. Second, one can reformulate \eqref{off-shell_scalar_1}-\eqref{off-shell_scalar_2}
as an unfolded Schwinger\textendash Dyson system and quantize the
theory this way.

\subsection{On-shell reduction: a scalar field of arbitrary mass\label{SEC_MASSIVE_SCALAR}}

We want to find an unfolded system that describes a scalar field $\phi(x)$
subjected to e.o.m.
\begin{equation}
(\square+2\lambda^{2}-m^{2})\phi=0.\label{massive_KG}
\end{equation}
This should arise as some reduction of the off-shell system \eqref{off-shell_scalar_1}-\eqref{off-shell_scalar_2}.
Looking at \eqref{box_scal_J} one sees that $g_{0}$ should be identified
with $m^{2}$, and $J_{0}^{(0)}(x)$ with $\phi(x)$. Then from \eqref{off-shell_scalar_2}
it follows that all $J^{(n)}(Y|x)$ and their equations have to represent
exact copies of $C(Y|x)$. Thus the submodule $J(Y|p|x)$ actually
disappears and one is left with an unfolded module $C(Y|x)$ subjected
to
\begin{equation}
DC+ie\partial\bar{\partial}C-iey\bar{y}C+iey\bar{y}\frac{(m/\lambda)^{2}}{(N+1)(N+2)}C=0.\label{unf_mass_scal_eq}
\end{equation}
This provides an unfolded formulation of the free on-shell scalar
field with mass $m$. To see this one acts on \eqref{unf_mass_scal_eq}
with an operator
\begin{equation}
D_{C,m}^{*}=-\frac{\lambda^{2}}{2}(D^{\alpha\dot{\beta}}-i\partial^{\alpha}\bar{\partial}^{\dot{\beta}}+iy^{\alpha}\bar{y}^{\dot{\beta}}-iy^{\alpha}\bar{y}^{\dot{\beta}}\frac{(m/\lambda)^{2}}{(N+1)(N+2)})\frac{\partial}{\partial e^{\alpha\dot{\beta}}},\label{D*_C_m}
\end{equation}
which is a generalization of \eqref{D*_C} for nonzero mass. It recovers
following wave equations from \eqref{unf_mass_scal_eq}
\begin{equation}
\square C_{N}+\lambda^{2}(N^{2}+2N+2)C_{N}-m^{2}C_{N}=0,
\end{equation}
so for a primary scalar $\phi(x)=C_{0}(0|x)$ one has \eqref{massive_KG}.

Also, one can define an unfolded system that corresponds to the \textquotedbl off-shell
scalar field with mass $m$\textquotedbl . By this we mean an unfolded
system which describes the coupling of the external current directly
to \eqref{massive_KG}. Although in terms of content it gives nothing
new compared to \eqref{off-shell_scalar_1}-\eqref{off-shell_scalar_2},
it is handy when dealing with the quantization problem considered
below. To construct this system one notices that from the standpoint
of formal consistency of \eqref{off-shell_scalar_1}-\eqref{off-shell_scalar_2}
the only requirement for $g\lambda^{-2}\frac{\partial}{\partial p}$
is to commute with all other operators in equations that act on $C$
and $J$. So if one shifts it by $(m/\lambda)^{2}$ the consistency
will be preserved. The resulting equations are
\begin{align}
 & DC+ie\partial\bar{\partial}C-iey\bar{y}C+iey\bar{y}\frac{(m/\lambda)^{2}}{(N+1)(N+2)}C=-iey\bar{y}\frac{g\lambda^{-2}}{(N+1)(N+2)}J^{(0)},\label{off-shell_mass_scalar_1}\\
 & DJ+ie\partial\bar{\partial}J-iey\bar{y}J+iey\bar{y}\frac{(m/\lambda)^{2}}{(N+1)(N+2)}J=-iey\bar{y}\frac{g\lambda^{-2}}{(N+1)(N+2)}\frac{\partial}{\partial p}J,\label{off-shell_mass_scalar_2}
\end{align}
and application of \eqref{D*_C_m} reveals a desired constraint
\begin{equation}
(\square+2\lambda^{2}-m^{2})\phi=gJ_{0}^{(0)}.
\end{equation}
One can also write down a more general off-shell system with varying
mass parameter: formally, one can take $m=m(p)$ being an arbitrary
function of $p$ in \eqref{off-shell_mass_scalar_2}. In this case
the current module represents an infinite sequence of glued $SO(3,2)$-modules
of scalar fields with arbitrary masses.

Finally, as was pointed out in \cite{misuna}, it is possible to impose
higher-order equations of motion. Consider for simplicity $m=0$ case
\eqref{off-shell_scalar_1}-\eqref{off-shell_scalar_2}. Then, if
one restricts the upper limit of $n$-summation in \eqref{scal_source_J}
with some $n_{0}$, resulting system will encode the following order-$(2n_{0}+4)$
equation of motion\footnote{If one starts from \eqref{off-shell_mass_scalar_1}-\eqref{off-shell_mass_scalar_2},
it is possible to get a general $(n_{0}+2)$-degree polynomial in
$\square$ with arbitrary coefficients on the l.h.s. of \eqref{high_order_eom}.}
\begin{equation}
(\square+2\lambda^{2})^{n_{0}+2}\phi(x)=0.\label{high_order_eom}
\end{equation}
In terms of the representation theory this restriction is tantamount
to the quotienting of the infinitely indecomposable off-shell scalar
Verma module by its submodule dual to \eqref{J_n0}. Resulting representation
is $(n_{0}+1)$ times indecomposable, with remaining $(n_{0}+1)$
nested submodules being dual to spaces of solutions to $(\square+2\lambda^{2})^{k}\phi=0$,
$k=[1,(n_{0}+1)]$. And the maximal on-shell reduction $J=0$, leading
to Klein\textendash Gordon equation \eqref{scal_eom}, implements
a quotient by the maximal submodule (dual to the whole $J$), which
results in an irreducible (after fixing boundary conditions) module
$D(1,0)\oplus D(2,0)$.

\subsection{Quantization: a scalar field propagator\label{SUBSEC_SCAL_QUANT}}

Here we present a sample calculation which illustrates how one can
calculate correlation functions from the off-shell unfolded system.
To this end we treat \eqref{off-shell_mass_scalar_1}-\eqref{off-shell_mass_scalar_2}
as Schwinger\textendash Dyson equations and try to solve them. As
we deal with a free theory with linear e.o.m. it is convenient to
introduce a connected generating functional $W=i\log Z$. Then a transition
from classical e.o.m. for field $\varphi_{n}$ to Schwinger\textendash Dyson
equation for $W$ is performed by a substitution $\varphi_{k}\rightarrow\frac{\delta W}{\delta J_{k}}$
and addition of $J_{n}$ to the r.h.s. Thus if we manage to solve
\eqref{off-shell_mass_scalar_1} for $J$ we will be able to restore
two-point function of a massive scalar field in $AdS_{4}$.

Applying \eqref{D*_C_m} to \eqref{off-shell_mass_scalar_1} leads
to
\begin{equation}
\left(\square+(N^{2}+2N+2)\lambda^{2}-m^{2}\right)C=\frac{g}{2(N+1)}\left(\frac{iy^{\alpha}\bar{y}^{\dot{\alpha}}}{(N+2)}D_{\alpha\dot{\alpha}}+N+2\right)J.\label{C_J_eq}
\end{equation}
This can be solved by means of a standard $AdS_{4}$ scalar propagator
$G(x,x')$ that solves
\begin{equation}
(\square-\mu^{2})G(x,x')=\delta(x,x').
\end{equation}
The propagator is expressed via the hypergeometric function as \cite{AdS_propagator}
\begin{equation}
G_{\Delta}(x,x')=\lambda^{2}A_{\Delta}\xi^{\Delta}F(\frac{\Delta}{2},\frac{\Delta+1}{2};\Delta-\frac{1}{2};\xi^{2}),\label{Green_function}
\end{equation}
where $\xi$ in the Poincaré coordinates is
\begin{equation}
\xi=\frac{2zz'}{z^{2}+z'^{2}+(\mathrm{x}-\mathrm{x}')^{2}},
\end{equation}
conformal weight is determined from
\begin{equation}
(\mu/\lambda)^{2}=\Delta(\Delta-3)
\end{equation}
and $A_{\Delta}$ is a $\Delta$-dependent normalization constant.
Thus for \eqref{C_J_eq} conformal weights are determined by
\begin{equation}
\Delta_{m,N}(\Delta_{m,N}-3)=(m/\lambda)^{2}-(N^{2}+2N+2),
\end{equation}
and a solution for \eqref{C_J_eq} is
\begin{equation}
C(Y|x)=\int dx'^{4}G_{\Delta_{m,N}}(x,x')\frac{g}{2(N+1)}\left(\frac{iy^{\alpha}\bar{y}^{\dot{\alpha}}}{(N+2)}D_{\alpha\dot{\alpha}}+N+2\right)J(Y|x').\label{C_(J)}
\end{equation}
This encodes two-point functions for all components of the unfolded
module $C(Y|x)$ in terms of the unfolded sources from $J(Y|x)$.
To extract them explicitly in terms of the primary scalar field $\phi(x)=C(0|x)$
one has to express descendants from $C(Y|x)$ in terms of $\phi$
and descendants from $J(Y|x)$ in terms of the primary source $J_{0}^{(0)}=J(Y=0|p=0|x)$.
Then by replacing $\phi\rightarrow\dfrac{\delta W}{\delta J_{0}^{(0)}}$
in \eqref{C_(J)} one can evaluate two-point functions. For instance,
for the propagator of the primary $\phi$ one has
\begin{equation}
\bigl\langle\phi(x)\phi(x')\bigr\rangle=gG_{\Delta_{m,0}}(x,x'),
\end{equation}
as, of course, it should be.

\section{Unfolded system for Fronsdal current\label{SEC_FRNSDL_CRRNT}}

Our goal is to apply the method of off-shell completion reasoned in
Section \ref{SEC_HS_UNFOLDED} and illustrated in Section \ref{SEC_SPIN_0}
to the Fronsdal field of arbitrary integer spin. To this end we should
unfold Fronsdal current \eqref{conserv_law} and then couple it to
the unfolded Fronsdal equations \eqref{OnShTh_1}-\eqref{OnShTh_2}.
This Section is devoted to the first part of the problem.

Double-traceless Fronsdal current $J_{\underline{a}(s)}^{F}(x)$ can
be decomposed into two traceless currents: $J_{\underline{a}(s)}(x)$
(traceless part of $J^{F})$ and $T_{\underline{a}(s-2)}(x)$ (trace
of $J^{F}$). Then the content of the generalized conservation law
\eqref{conserv_law} is that the divergence $D^{\underline{b}}J_{\underline{b}\underline{a}(s-1)}$
is proportional to the first symmetrized derivative $D_{\underline{a}}T_{\underline{a}(s-2)}$.
Apart from this, $J_{\underline{a}(s)}$ and $T_{\underline{a}(s-2)}$
are unconstrained. So we unfold the Fronsdal current in the following
way: first, in Subsection \ref{SEC_J} we unfold traceless conserved
$J_{\underline{a}(s)}$; based on this, in Subsection \ref{SEC_T}
we find an unfolded formulation for unconstrained traceless $T_{\underline{a}(s-2)}$;
finally, in Subsection \ref{SEC_J_T} we couple $T$ to $J$ in such
a way as to impose \eqref{conserv_law}, which completes the procedure.
Also, as we show in Subsection \ref{SUBSEC_FIERZ-PAULI}, the unfolded
system for conserved $J$ admits a reduction to an unfolded system
that describes on-shell massive HS fields subjected to Fierz-Pauli
conditions. 

\subsection{Conserved traceless HS current\label{SEC_J}}

The first task is to find an unfolded system for traceless spin-$s$
current $J_{\underline{a}(s)}(x)$ subjected to the conservation condition
\begin{equation}
D^{\underline{b}}J_{\underline{ba}(s-1)}=0.\label{conserv_J}
\end{equation}
We solve the problem in two steps. First, we construct an unfolded
system for $J_{\underline{a}(s)}$ that in addition to \eqref{conserv_J}
also satisfies \textquotedbl masslessness\textquotedbl{} condition
\begin{equation}
\square J_{\underline{a}(s)}+2\lambda^{2}J_{\underline{a}(s)}=0.\label{box_J_0}
\end{equation}
Then we remove \eqref{box_J_0} by introducing additional descendants,
similar to what we did for the scalar field.

An appropriate unfolded module for \eqref{conserv_J}, \eqref{box_J_0}
is $J(Y|x)$ such that
\begin{equation}
\varsigma\geq s,\quad|\tau|\leq s.\label{s_t_limit}
\end{equation}
In Lorentz tensor language this module corresponds to the space of
all one- and two-row traceless Young diagrams with the upper (the
only for one-row diagrams) row length at least $s$ and lower row
length at most $s$. This is because the primary source $J_{\underline{a}(s)}$
represents a Young diagram with one row of length $s$, and successive
differentiation of $J_{\underline{a}(s)}$ will add new cells to the
diagram, corresponding to the traceless-symmetrized derivatives. All
contractions or antisymmetrizations of derivatives give nothing new
due to \eqref{conserv_J}, \eqref{box_J_0} and the commutator of
$AdS$-covariant derivatives.

The most general Ansatz for an unfolded equation is
\begin{equation}
DJ+ie\partial\bar{\partial}a_{N,\bar{N}}J-iey\bar{y}b_{N,\bar{N}}J+ey\bar{\partial}c_{N,\bar{N}}J+e\bar{y}\partial\bar{c}_{N,\bar{N}}J=0.
\end{equation}
Coefficients $a_{N,\bar{N}}$, $b_{N,\bar{N}}$, $c_{N,\bar{N}}$
and $\bar{c}_{N,\bar{N}}$ are (partially) fixed by the consistency
requirement \eqref{unf_consist}. Imposing it one arrives at the following
recurrent system
\begin{align}
 & (N+2)a_{N+1,\bar{N}-1}c_{N,\bar{N}}-Na_{N,\bar{N}}c_{N-1,\bar{N}-1}=0,\label{rec3}\\
 & Nb_{N-1,\bar{N}+1}\bar{c}_{N,\bar{N}}-(N+2)b_{N,\bar{N}}\bar{c}_{N+1,\bar{N}+1}=0,\label{rec5}\\
 & (N+2)a_{N+1,\bar{N}+1}b_{N,\bar{N}}-Na_{N,\bar{N}}b_{N-1,\bar{N}-1}+(N+2)c_{N,\bar{N}}\bar{c}_{N+1,\bar{N}-1}-N\bar{c}_{N,\bar{N}}c_{N-1,\bar{N}+1}=2,\label{rec1}
\end{align}
plus conjugate equations resulting from $N\leftrightarrow\bar{N}$,
$a_{K,\bar{L}}\rightarrow a_{L,\bar{K}}$, $b_{K,\bar{L}}\rightarrow b_{L,\bar{K}}$
and $c_{K,\bar{L}}\leftrightarrow\bar{c}_{L,\bar{K}}$.

As mentioned previously, partial leftover freedom in fixing unfolded
coefficients helps to simplify the system, but may affect further
nonlinear deformations or couplings to other systems. In fact, such
a simplification is even necessary to some extent, because it seems
unfeasible to explicitly solve all consistency conditions in their
full generality. The final goal of the current analysis is to couple
an unfolded module $J(Y|x)$ to the unfolded Fronsdal equations \eqref{OnShTh_1}-\eqref{OnShTh_2}
with $J_{\alpha(s),\dot{\alpha}(s)}(x)$ playing the role of the traceless
part of the Fronsdal spin-$s$ current. It turns out that not every
possible solution of \eqref{rec3}-\eqref{rec1} leads to this result.

The proper choice is to put
\begin{align}
 & a_{N,\bar{N}}=1,\quad\varsigma>s,\label{a_1}\\
 & a_{N,\bar{N}}=0,\quad\varsigma=s,\label{a_1_bndr}
\end{align}
where \textquotedbl boundary condition\textquotedbl{} \eqref{a_1_bndr}
is dictated by self-consistency: nonzero $a_{N,\bar{N}}$ with $\varsigma=s$
would give rise to descendants with tensor rank $(s-1)$ which are
absent in $J$ by construction. A different solution, which has somewhat
simpler-looking coefficients but does not allow the identification
of $J_{\alpha(s),\dot{\alpha}(s)}$ with a traceless part of Fronsdal
current, is given in Appendix B.

Applying \eqref{a_1} to \eqref{rec3}-\eqref{rec5} and their conjugate
allows one to solve for $b_{N,\bar{N}}$, $c_{N,\bar{N}}$ and $\bar{c}_{N,\bar{N}}$
\begin{align}
 & b_{N,\bar{N}}=\frac{b_{\varsigma}}{(N+1)(N+2)(\bar{N}+1)(\bar{N}+2)},\label{b_part}\\
 & c_{N,\bar{N}}=\frac{c_{\tau}}{(N+1)(N+2)},\label{c_part}\\
 & \bar{c}_{N,\bar{N}}=\frac{\bar{c}_{\tau}}{(\bar{N}+1)(\bar{N}+2)},\label{c-_part}
\end{align}
where $b_{\varsigma}$ and $c_{\tau}$, $\bar{c}_{\tau}$ are so far
arbitrary functions of $\varsigma$ and $\tau$ respectively. To determine
them one adds to \eqref{rec1} its conjugate and substituting \eqref{b_part}-\eqref{c_part}
arrives at
\begin{equation}
2\left((\varsigma+1)^{4}-(\varsigma+1)^{2}-\tau^{4}+\tau^{2}\right)+(\varsigma+2)b_{\varsigma-1}-\varsigma b_{\varsigma}+(\tau-1)c_{\tau}\bar{c}_{\tau+1}-(\tau+1)\bar{c}_{\tau}c_{\tau-1}=0.
\end{equation}
From here one finds, separating $\varsigma$ and $\tau$ variables,
\begin{align}
 & b_{\varsigma}=(\varsigma+1)^{2}(\varsigma+2)^{2}+(\varsigma+1)(\varsigma+2)b,\\
 & \bar{c}_{\tau}c_{\tau-1}=\tau^{2}(\tau-1)^{2}+\tau(\tau-1)c,
\end{align}
with $b$ and $c$ being arbitrary constants. To fix them one notices
that restrictions \eqref{s_t_limit} put following \textquotedbl boundary
conditions\textquotedbl{} on $b_{\varsigma}$, $c_{\tau}$ and $\bar{c}_{\tau}$
\begin{equation}
b_{\varsigma=s-1}=0,\quad c_{\tau=s}=0,\quad\bar{c}_{\tau=-s}=0.
\end{equation}
To satisfy them one chooses
\begin{equation}
b=c=-s(s+1).
\end{equation}
Then an unfolded system for the traceless spin-$s$ current $J_{\underline{a}(s)}$
subjected to \eqref{conserv_J} and \eqref{box_J_0} is
\begin{align}
 & DJ+ie\partial\bar{\partial}J-iey\bar{y}\frac{(\varsigma+1)(\varsigma+2)(\varsigma-s+1)(\varsigma+s+2)}{(N+1)(N+2)(\bar{N}+1)(\bar{N}+2)}J+\nonumber \\
 & +ey\bar{\partial}\frac{\tau(\tau-s)}{(N+1)(N+2)}J+e\bar{y}\partial\frac{\tau(\tau+s)}{(\bar{N}+1)(\bar{N}+2)}J=0.\label{massless_J}
\end{align}

Now one has to relax masslessness condition \eqref{box_J_0}. The
idea is the same as for the scalar field: one introduces an infinite
sequence of descendant modules $J^{(n)}(Y|x)$, successively contributing
to the r.h.s. of \eqref{box_J_0} for each other, and collects them
into single $J(Y|p|x)=\sum_{n=0}^{\infty}\frac{p^{n}}{n!}J^{(n)}(Y|x)$.
So one seeks a generalization of \eqref{massless_J} of the form
\begin{align}
 & DJ+ie\partial\bar{\partial}J-iey\bar{y}\frac{(\varsigma+1)(\varsigma+2)(\varsigma-s+1)(\varsigma+s+2)}{(N+1)(N+2)(\bar{N}+1)(\bar{N}+2)}J+\nonumber \\
 & +ey\bar{\partial}\frac{\tau(\tau-s)}{(N+1)(N+2)}J+e\bar{y}\partial\frac{\tau(\tau+s)}{(\bar{N}+1)(\bar{N}+2)}J+\nonumber \\
 & +ie\partial\bar{\partial}\hat{a}_{N,\bar{N}}\frac{\partial}{\partial p}J+iey\bar{y}\hat{b}_{N,\bar{N}}\frac{\partial}{\partial p}J+ey\bar{\partial}\hat{c}_{N,\bar{N}}\frac{\partial}{\partial p}J+e\bar{y}\partial\hat{\bar{c}}_{N,\bar{N}}\frac{\partial}{\partial p}J=0,\label{J_p_ansatz}
\end{align}
with some coefficients $\hat{a}_{N,\bar{N}}$, $\hat{b}_{N,\bar{N}}$,
$\hat{c}_{N,\bar{N}}$ and $\hat{\bar{c}}_{N,\bar{N}}$ . In principle,
one could allow these coefficients to depend on $p$, but this is
an unnecessary overgeneralization, as follows from the isomorphism
of all $J^{(n)}$.

Once again, it is possible to fix a part of unknown coefficients from
the very beginning. New descendants $J^{(n)}$ encode d'Alembertians
of the primary source so, from the standpoint of Young diagrams, these
descendants arise from removal of cells. So we keep only those terms
that correspond to such removing and take
\begin{equation}
\hat{a}_{N,\bar{N}}=0,\quad\hat{c}_{N,\bar{N}}|_{\tau<0}=0,\quad\hat{\bar{c}}_{N,\bar{N}}|_{\tau>0}=0.\label{vanish_a_c_c-}
\end{equation}
Then formal consistency \eqref{unf_consist} requires
\begin{align}
 & (N+2)\hat{c}_{N,\bar{N}}-N\hat{c}_{N-1,\bar{N}-1}=0,\label{rec7}\\
 & \frac{(\varsigma+1)(\varsigma+2)(\varsigma-s+1)(\varsigma+s+2)}{(N+1)(\bar{N}+3)}\hat{\bar{c}}_{N,\bar{N}}+(N+2)\frac{\tau(\tau+s)}{(\bar{N}+3)}\hat{b}_{N,\bar{N}}-\nonumber \\
 & -\frac{(\varsigma+1)(\varsigma+2)(\varsigma-s+1)(\varsigma+s+2)}{(N+1)(\bar{N}+1)}\hat{\bar{c}}_{N+1,\bar{N}+1}-N\frac{\tau(\tau+s)}{(\bar{N}+1)}\hat{b}_{N-1,\bar{N}+1}=0.\label{rec10}\\
 & N\hat{b}_{N-1,\bar{N}-1}+\frac{\tau(\tau-s)}{(N+1)}\hat{\bar{c}}_{N+1,\bar{N}-1}+(N+2)\frac{(\tau+1)(\tau+s+1)}{\bar{N}(\bar{N}+1)}\hat{c}_{N,\bar{N}}-\nonumber \\
 & -(N+2)\hat{b}_{N,\bar{N}}-\frac{(\tau-1)(\tau-s-1)}{(N+1)}\hat{\bar{c}}_{N,\bar{N}}-N\frac{\tau(\tau+s)}{(\bar{N}+1)(\bar{N}+2)}\hat{c}_{N-1,\bar{N}+1}=0,\label{rec12}\\
 & (N+2)\hat{b}_{N,\bar{N}}\hat{\bar{c}}_{N+1,\bar{N}+1}-N\hat{b}_{N-1,\bar{N}+1}\hat{\bar{c}}_{N,\bar{N}}=0,\label{rec14}\\
 & (N+2)\hat{\bar{c}}_{N+1,\bar{N}-1}\hat{c}_{N,\bar{N}}-N\hat{c}_{N-1,\bar{N}+1}\hat{\bar{c}}_{N,\bar{N}}=0,\label{rec16}
\end{align}
plus conjugate equations. From \eqref{rec7} and conjugate one finds
\begin{align}
 & \hat{c}_{N,\bar{N}}|_{\tau\geq0}=\frac{\hat{c}_{\tau}}{(N+1)(N+2)},\label{c_hat_part}\\
 & \hat{\bar{c}}_{N,\bar{N}}|_{\tau\leq0}=\frac{\hat{\bar{c}}_{\tau}}{(\bar{N}+1)(\bar{N}+2)},\label{c-_hat_part}
\end{align}
with $\hat{c}_{\tau}$ being arbitrary function of $\tau$ and $\hat{\bar{c}}_{\tau}$
being its conjugate $\hat{\bar{c}}_{\tau}=\hat{c}_{-\tau}$ in order
to ensure reality. After substituting \eqref{c-_hat_part} to \eqref{rec10}
terms with $\hat{\bar{c}}_{\tau}$ cancel and for $\hat{b}_{N,\bar{N}}$
one obtains
\begin{equation}
\hat{b}_{N,\bar{N}}=\frac{\hat{b}_{\varsigma}}{(N+1)(N+2)(\bar{N}+1)(\bar{N}+2)}
\end{equation}
with arbitrary $\varsigma$-dependent $\hat{b}_{\varsigma}$. Now,
the sum and the difference of \eqref{rec12} and its conjugate yield,
respectively,
\begin{align}
 & (\tau+1)\left((\tau-1)(\tau-1-s)\hat{\bar{c}}_{\tau}+\tau(\tau+s)\hat{c}_{\tau-1}\right)+(\varsigma+2)\hat{b}_{\varsigma-1}-\nonumber \\
 & -(\tau-1)\left((\tau+1)(\tau+1+s)\hat{c}_{\tau}+\tau(\tau-s)\hat{\bar{c}}_{\tau+1}\right)-\varsigma\hat{b}_{\varsigma}=0.\\
 & (\varsigma+1)\left((\tau+1)(\tau+1+s)\hat{c}_{\tau}+\tau(\tau-s)\hat{\bar{c}}_{\tau+1}\right)+\tau\hat{b}_{\varsigma}-\nonumber \\
 & -(\varsigma+1)\left((\tau-1)(\tau-1-s)\hat{\bar{c}}_{\tau}+\tau(\tau+s)\hat{c}_{\tau-1}\right)-\tau\hat{b}_{\varsigma-1}=0,
\end{align}
Separating $\varsigma$ and $\tau$ variables and making use of the
reality condition $\hat{c}_{\tau}=\hat{\bar{c}}_{-\tau}$ one easily
finds, taking into account \eqref{vanish_a_c_c-},
\begin{align}
 & \hat{b}_{\varsigma}=\hat{b}+\hat{c}(\varsigma+1)(\varsigma+2),\\
 & \hat{c}_{\tau\geq0}=-\frac{\hat{b}+\hat{c}\tau(\tau+1)}{(\tau+1)(\tau+1+s)},\\
 & \hat{\bar{c}}_{\tau\leq0}=-\frac{\hat{b}+\hat{c}\tau(\tau-1)}{(\tau-1)(\tau-1-s)},
\end{align}
with $\hat{b}$ and $\hat{c}$ being arbitrary constants. Restrictions
on $\varsigma$ and $\tau$ values \eqref{s_t_limit} require $\hat{\bar{c}}_{-s}=0$
and $\hat{c}_{s}=0$, which entail
\begin{equation}
\hat{b}=-\hat{c}s(s+1).
\end{equation}
Then by rescaling $p$ in \eqref{J_p_ansatz} one can always set
\begin{equation}
\hat{c}=1,\quad\hat{b}=-s(s+1).
\end{equation}
Remaining equations \eqref{rec14}, \eqref{rec16} and their conjugate
now hold identically and give no further constraints.

So, now one is in a position to write down an unfolded system that
describes a conserved traceless spin-$s$ current. It is handy to
introduce projectors $\Pi^{+}$ and $\Pi^{-}$ to components with
$\tau\geq0$ and $\tau\leq0$
\begin{equation}
\Pi^{+}F_{N,\bar{N}}(Y)=\begin{cases}
F_{N,\bar{N}}(Y), & N\geq\bar{N}\\
0, & N<\bar{N}
\end{cases};\qquad\Pi^{-}F_{N,\bar{N}}(Y)=\begin{cases}
F_{N,\bar{N}}(Y), & N\leq\bar{N}\\
0, & N>\bar{N}
\end{cases}.
\end{equation}
Then the system in question is realized by an unfolded module $J(Y|p|x)$
with $\varsigma\geq s$, $|\tau|\leq s$ subjected to unfolded equation
\begin{align}
 & DJ+ie\partial\bar{\partial}J-iey\bar{y}\frac{(\varsigma+1)(\varsigma+2)(\varsigma-s+1)(\varsigma+s+2)}{(N+1)(N+2)(\bar{N}+1)(\bar{N}+2)}J+\nonumber \\
 & +ey\bar{\partial}\frac{\tau(\tau-s)}{(N+1)(N+2)}J+e\bar{y}\partial\frac{\tau(\tau+s)}{(\bar{N}+1)(\bar{N}+2)}J+iey\bar{y}\frac{(\varsigma-s+1)(\varsigma+s+2)}{(N+1)(N+2)(\bar{N}+1)(\bar{N}+2)}\frac{\partial}{\partial p}J-\nonumber \\
 & -ey\bar{\partial}\frac{(\tau-s)}{(\tau+1)(N+1)(N+2)}\frac{\partial}{\partial p}\Pi^{+}J-e\bar{y}\partial\frac{(\tau+s)}{(\tau-1)(\bar{N}+1)(\bar{N}+2)}\frac{\partial}{\partial p}\Pi^{-}J=0.\label{J_conserved_eq}
\end{align}

\subsubsection{On-shell reduction: Fierz-Pauli massive HS fields\label{SUBSEC_FIERZ-PAULI}}

Interestingly, the unfolded formulation for conserved spin-$s$ current
we found also allows for a description of the on-shell massive HS
fields. Concretely, equation \eqref{J_conserved_eq} admits a simple
reduction that sends it to the unfolded system, corresponding to the
($AdS$-generalization of) Fierz-Pauli equations for a massive spin-$s$
field
\begin{align}
 & (\square-m^{2})\phi_{\underline{a}(s)}=0,\label{FP1}\\
 & D^{\underline{b}}\phi_{\underline{ba}(s-1)}=0,\\
 & \phi^{\underline{b}}\text{}_{\underline{ba}(s-2)}=0.\label{FP3}
\end{align}
A general idea is the same as was used for the scalar field in Subsection
\ref{SEC_MASSIVE_SCALAR}: from the standpoint of formal consistency
of \eqref{J_conserved_eq}, the only requirement for $\frac{\partial}{\partial p}$
is to commute with all other operators in the equation. So one can
reduce the unfolded module $J(Y|p|x)\rightarrow J(Y|x)$ replacing
$\frac{\partial}{\partial p}$ in \eqref{J_conserved_eq} with $(m/\lambda)^{2}$
and this will not ruin the consistency. Then one arrives at
\begin{align}
 & DJ+ie\partial\bar{\partial}J-iey\bar{y}\frac{(\varsigma-s+1)(\varsigma+s+2)\left((\varsigma+1)(\varsigma+2)-(m/\lambda)^{2}\right)}{(N+1)(N+2)(\bar{N}+1)(\bar{N}+2)}J+\nonumber \\
 & +ey\bar{\partial}\frac{(\tau-s)}{(N+1)(N+2)}\left(\tau-\frac{(m/\lambda)^{2}}{(\tau+1)}\Pi^{+}\right)J+e\bar{y}\partial\frac{(\tau+s)}{(\bar{N}+1)(\bar{N}+2)}\left(\tau-\frac{(m/\lambda)^{2}}{(\tau-1)}\Pi^{-}\right)J=0.\label{Fierz-Pauli}
\end{align}
Acting with a \textquotedbl conjugate operator\textquotedbl
\begin{align}
 & D_{FP}^{*}=-\frac{\lambda^{2}}{2}\Biggl(D^{\alpha\dot{\beta}}-i\partial^{\alpha}\bar{\partial}^{\dot{\beta}}+iy^{\alpha}\bar{y}^{\dot{\beta}}\frac{(\varsigma-s+1)(\varsigma+s+2)\left((\varsigma+1)(\varsigma+2)-(m/\lambda)^{2}\right)}{(N+1)(N+2)(\bar{N}+1)(\bar{N}+2)}-\nonumber \\
 & -y^{\alpha}\bar{\partial}^{\dot{\beta}}\frac{(\tau-s)}{(N+1)(N+2)}\left(\tau-\frac{(m/\lambda)^{2}}{(\tau+1)}\Pi^{+}\right)J-\bar{y}^{\dot{\beta}}\partial^{\alpha}\frac{(\tau+s)}{(\bar{N}+1)(\bar{N}+2)}\left(\tau-\frac{(m/\lambda)^{2}}{(\tau-1)}\Pi^{-}\right)\Biggr)\frac{\partial}{\partial e^{\alpha\dot{\beta}}}\label{D*_FP}
\end{align}
on \eqref{Fierz-Pauli}, one reveals the following wave equation for
$J$
\begin{equation}
\square J-m^{2}J+\lambda^{2}(\varsigma(\varsigma+1)+\tau^{2}-s(s+1)+2)J=0.
\end{equation}
For the primary field $J_{\alpha(s),\dot{\alpha}(s)}(x)$ this gives
\begin{equation}
(\square-m^{2}+2\lambda^{2})J_{\alpha(s),\dot{\alpha}(s)}=0.
\end{equation}
Tracelessness and conservation condition \eqref{conserv_J} were built-in
from the very beginning. Thus \eqref{Fierz-Pauli} indeed provides
an unfolded formulation for Fierz-Pauli massive HS fields. \eqref{Fierz-Pauli}
also allows for an immediate off-shell completion: one just has to
restore $p$-dependence of the module $J(Y|p|x)$ and add $\frac{\partial}{\partial p}$
to each $(m/\lambda)^{2}$ in \eqref{Fierz-Pauli}. Resulting system
for \textquotedbl off-shell HS field with mass $m$\textquotedbl{}
will be equivalent to \eqref{J_conserved_eq} up to field redefinitions
(and will coincide literally in $m=0$ case), like it was for \textquotedbl off-shell
scalar field with mass $m$\textquotedbl{} in Subsection \ref{SEC_MASSIVE_SCALAR}.

\subsection{Trace of the Fronsdal current\label{SEC_T}}

The next task is to build an unfolded system for the trace of spin-$s$
Fronsdal current. This represents a rank-$(s-2)$ symmetric traceless
tensor $T_{\underline{a}(s-2)}(x)$ whose first symmetrized derivative
is proportional to the divergence of $J$ according to \eqref{conserv_law}.
So we are going to unfold an unconstrained symmetric traceless rank-$(s-2)$
tensor and then add $T$-dependent terms to \eqref{J_conserved_eq}.

The distinction between $T$-module and $J$-module studied before
is that now a primary field possesses unconstrained divergences, which
generate unfolded submodules corresponding to lower-rank tensors.
In order to account for them we introduce another expansion parameter
$q$ in addition to $p$
\begin{equation}
T(Y|p,q|x)=\sum_{n=0}^{\infty}\sum_{m=0}^{s-2}\frac{p^{n}}{n!}\frac{q^{m}}{m!}T^{(n,m)}(Y|x).
\end{equation}
Submodules $T^{(n,m)}$, corresponding to higher $q$-powers $m$,
contain order-$m$ divergences of the submodules $T^{(n,0)}$, so
in each $T^{(n,m)}$ a primary field is a tensor of rank $(s-m-2)$.
Thus instead of \eqref{s_t_limit} now one has for $T(Y|p,q|x)$
\begin{equation}
\varsigma\geq s-2-\nu,\quad|\tau|\leq s-2-\nu,\label{s_t_limit_T}
\end{equation}
where a $q$-power counting operator $\nu$ is defined as
\begin{equation}
\nu=q\frac{\partial}{\partial q}.
\end{equation}
Then an appropriate Ansatz for the unfolded system in question is
\begin{align}
 & DT+ie\partial\bar{\partial}T-iey\bar{y}\frac{(\varsigma-s+\nu+3)(\varsigma+s-\nu)}{(N+1)(N+2)(\bar{N}+1)(\bar{N}+2)}\left((\varsigma+1)(\varsigma+2)-\frac{\partial}{\partial p}-\mu_{\nu}\right)T+\nonumber \\
 & +e\bar{y}\partial\frac{\tau(\tau+s-\nu-2)}{(\bar{N}+1)(\bar{N}+2)}T+ey\bar{\partial}\frac{\tau(\tau-s+\nu+2)}{(N+1)(N+2)}T-\nonumber \\
 & -ey\bar{\partial}\frac{(\tau-s+\nu+2)}{(\tau+1)(N+1)(N+2)}(\frac{\partial}{\partial p}+\mu_{\nu})\Pi^{+}T-e\bar{y}\partial\frac{(\tau+s-\nu-2)}{(\tau-1)(\bar{N}+1)(\bar{N}+2)}(\frac{\partial}{\partial p}+\mu_{\nu})\Pi^{-}T+\nonumber \\
 & +iey\bar{y}\tilde{b}_{N,\bar{N}}^{\nu}\dfrac{\partial}{\partial q}T-ey\bar{\partial}\tilde{c}_{N,\bar{N}}^{\nu}\Pi^{+}\dfrac{\partial}{\partial q}T-e\bar{y}\partial\tilde{\bar{c}}_{N,\bar{N}}^{\nu}\Pi^{-}\dfrac{\partial}{\partial q}T=0.\label{T_Ansatz}
\end{align}
This Ansatz represents a generalization of \eqref{J_conserved_eq}
(for a $(s-2)$-rank tensor) which now includes the terms encoding
divergences (the last line) and accounts for reducing of the tensor
rank of divergences (so all $s$ have been replaced with $(s-\nu-2)$
as compared to \eqref{J_conserved_eq}). We also allowed for mass-like
terms (reproducing those from Subsection \ref{SUBSEC_FIERZ-PAULI})
with $\nu$-dependent parameter $\mu_{\nu}$. Motivation for introducing
them is that divergences of massless $AdS$-field of high spin have
conformal dimensions different from dimensions of massless fields
of lower spins (e.g. the dimension of a massless spin-1 divergence
$D_{\underline{n}}A^{\underline{n}}(x)$ does not coincide with that
of massless scalar $\phi$) which fact should reveal itself as emergence
of effective \textquotedbl masses\textquotedbl{} for the lower-rank
descendant modules. To put it differently, in $AdS$ space the wave
operator does not commute with covariant derivatives, therefore descendants
obey different wave equations than the primary.

Now one has to solve for the consistency conditions in order to fix
coefficients $\tilde{b}_{N,\bar{N}}^{\nu}$, $\tilde{c}_{N,\bar{N}}^{\nu}$,
$\tilde{\bar{c}}_{N,\bar{N}}^{\nu}$ and $\mu_{\nu}$. These comprise
following equations (and their conjugate)
\begin{align}
 & (N+2)\tilde{c}_{N,\bar{N}}^{\nu}\Pi^{+}-N\tilde{c}_{N-1,\bar{N}-1}^{\nu}\Pi^{+}=0,\label{rec_1}\\
 & \frac{(\bar{N}+2)\tau(\tau-s+\nu+2)}{(N+3)}\tilde{b}_{N,\bar{N}}^{\nu}-\bar{N}\frac{\tau(\tau-s+\nu+3)}{(N+1)}\tilde{b}_{N+1,\bar{N}-1}^{\nu}+\nonumber \\
 & +\frac{(\varsigma-s+\nu+4)(\varsigma+s-\nu-1)}{(N+1)(\bar{N}+1)}\left((\varsigma+1)(\varsigma+2)-\mu_{\nu-1}\right)\tilde{c}_{N+1,\bar{N}+1}^{\nu}\Pi^{+}-\nonumber \\
 & -\frac{(\varsigma-s+\nu+3)(\varsigma+s-\nu)}{(N+3)(\bar{N}+1)}\left((\varsigma+1)(\varsigma+2)-\mu_{\nu}\right)\tilde{c}_{N,\bar{N}}^{\nu}\Pi^{+}+\nonumber \\
 & +\frac{\bar{N}(\tau-s+3+\nu)\mu_{\nu-1}}{(\tau+1)(N+1)}\tilde{b}_{N+1,\bar{N}-1}^{\nu}\Pi^{+}-\frac{(\bar{N}+2)(\tau-s+2+\nu)\mu_{\nu}}{(\tau+1)(N+3)}\tilde{b}_{N,\bar{N}}^{\nu}\Pi^{+}=0,\label{rec_2}
\end{align}
\begin{align}
 & \frac{(\varsigma-s+3+\nu)(\varsigma+s-\nu)}{\left(N+3\right)\left(\bar{N}+1\right)}\tilde{c}_{N,\bar{N}}^{\nu}\Pi^{+}-\frac{(\varsigma-s+4+\nu)(\varsigma+s-\nu-1)}{(N+1)\left(\bar{N}+1\right)}\tilde{c}_{N+1,\bar{N}+1}^{\nu}\Pi^{+}+\nonumber \\
 & +\bar{N}\frac{(\tau-s+3+\nu)}{\left(\tau+1\right)\left(N+1\right)}\tilde{b}_{N+1,\bar{N}-1}^{\nu}\Pi^{+}-(\bar{N}+2)\frac{(\tau-s+2+\nu)}{\left(\tau+1\right)\left(N+3\right)}\tilde{b}_{N,\bar{N}}^{\nu}\Pi^{+}=0,\label{rec_3}\\
 & \bar{N}\tilde{b}_{N-1,\bar{N}-1}^{\nu}-(\bar{N}+2)\tilde{b}_{N,\bar{N}}^{\nu}+\frac{(\tau+1)(\tau-1+s-\nu)}{(\bar{N}+1)}\tilde{c}_{N,\bar{N}}^{\nu}\Pi^{+}-\frac{\tau(\tau+s-\nu-3)}{(\bar{N}+1)}\tilde{c}_{N-1,\bar{N}+1}^{\nu}\Pi^{+}+\nonumber \\
 & +\bar{N}\frac{\tau(\tau-s+\nu+3)}{(N+1)(N+2)}\tilde{\bar{c}}_{N+1,\bar{N}-1}^{\nu}\Pi^{-}-(\bar{N}+2)\frac{(\tau-1)(\tau+1-s+\nu)}{N(N+1)}\tilde{\bar{c}}_{N,\bar{N}}^{\nu}\Pi^{-}=0,\label{rec_4}\\
 & \bar{N}\tilde{b}_{N+1,\bar{N}-1}^{\nu}\tilde{c}_{N,\bar{N}}^{\nu+1}\Pi^{+}-(\bar{N}+2)\tilde{b}_{N,\bar{N}}^{\nu+1}\tilde{c}_{N+1,\bar{N}+1}^{\nu}\Pi^{+}=0.\label{rec_5}
\end{align}
From \eqref{rec_1} it follows that
\begin{equation}
\tilde{c}_{N,\bar{N}}^{\nu}=\frac{\tilde{c}_{\tau}^{\nu}}{(N+1)(N+2)}.
\end{equation}
Considering \eqref{rec_2} with $\tau<0$ excludes all terms with
$\Pi^{+}$ and allows to solve for negative-$\tau$ $\tilde{b}_{N,\bar{N}}^{\nu}$.
Combining this with positive-$\tau$ answer resulting from conjugate
equation one can write
\begin{equation}
\tilde{b}_{N,\bar{N}}^{\nu}=\frac{\tilde{b}_{\varsigma}^{\nu}}{(|\tau|+s-\nu-2)(N+1)(N+2)(\bar{N}+1)(\bar{N}+2)}.\label{b_tilde}
\end{equation}
Plugging this into \eqref{rec_3} leads to
\begin{equation}
\tilde{b}_{N,\bar{N}}^{\nu}=\frac{\tilde{b}^{\nu}}{(|\tau|+s-\nu-2)(N+1)(N+2)(\bar{N}+1)(\bar{N}+2)},\label{b_tilde_nu}
\end{equation}
\begin{equation}
\tilde{c}_{N,\bar{N}}^{\nu}=\frac{\tilde{b}^{\nu}}{(\tau+1)(\tau+s-\nu-2)(\tau+s-\nu-1)(N+1)(N+2)},\label{c_tilde}
\end{equation}
and by conjugacy
\begin{equation}
\tilde{\bar{c}}_{N,\bar{N}}^{\nu}=\frac{\tilde{b}^{\nu}}{(|\tau|+1)(|\tau|+s-\nu-2)(|\tau|+s-\nu-1)(\bar{N}+1)(\bar{N}+2)}.\label{c_tilde_bar_nu}
\end{equation}
$\tilde{b}^{\nu}$ in \eqref{b_tilde_nu}-\eqref{c_tilde_bar_nu}
is an arbitrary function of $\nu$. One can always set it to one
\begin{equation}
\tilde{b}^{\nu}=1
\end{equation}
by rescaling the $q$ variable. Finally, returning to \eqref{rec_2}
one now determines $\mu_{\nu}$ to be
\begin{equation}
\mu_{\nu}=\mu-\nu(2s-\nu-3)\label{mu}
\end{equation}
with arbitrary constant $\mu$. The rest of equations \eqref{rec_4},
\eqref{rec_5} and conjugate then hold identically. Thus, unfolded
system \eqref{T_Ansatz} with coefficients \eqref{b_tilde_nu}-\eqref{mu}
describes unrestricted symmetric traceless rank-$(s-2)$ tensor field.

\subsection{Double-traceless Fronsdal current\label{SEC_J_T}}

Now to build an unfolded system for Fronsdal current one adds $T$-dependent
terms to \eqref{J_conserved_eq} in such a way that \eqref{conserv_J}
for the primary of the unfolded $J$-module gets deformed to \eqref{conserv_law}
with $J_{\alpha(s),\dot{\alpha}(s)}$ and $T_{\alpha(s-2),\dot{\alpha}(s-2)}$
being traceless and trace parts of $J_{\underline{a}(s)}^{F}$. According
to \eqref{conserv_law} $T$ must couple as a divergence of $J$,
so one has the following Ansatz
\begin{align}
 & DJ+ie\partial\bar{\partial}J-iey\bar{y}\frac{(\varsigma+1)(\varsigma+2)(\varsigma-s+1)(\varsigma+s+2)}{(N+1)(N+2)(\bar{N}+1)(\bar{N}+2)}J+\nonumber \\
 & +e\bar{y}\partial\frac{\tau(\tau+s)}{(\bar{N}+1)(\bar{N}+2)}J+ey\bar{\partial}\frac{\tau(\tau-s)}{(N+1)(N+2)}J+iey\bar{y}\frac{(\varsigma-s+1)(\varsigma+s+2)}{(N+1)(N+2)(\bar{N}+1)(\bar{N}+2)}\frac{\partial}{\partial p}J-\nonumber \\
 & -ey\bar{\partial}\frac{(\tau-s)}{(\tau+1)(N+1)(N+2)}\frac{\partial}{\partial p}\Pi^{+}J-e\bar{y}\partial\frac{(\tau+s)}{(\tau-1)(\bar{N}+1)(\bar{N}+2)}\frac{\partial}{\partial p}\Pi^{-}J+\nonumber \\
 & +iey\bar{y}\frac{\check{b}_{\tau}}{(N+1)(N+2)(\bar{N}+1)(\bar{N}+2)}T-ey\bar{\partial}\frac{\check{c}_{\tau}}{(N+1)(N+2)}\Pi^{+}T-e\bar{y}\partial\frac{\check{\bar{c}}_{\tau}}{(\bar{N}+1)(\bar{N}+2)}\Pi^{-}T=0.\label{J_eq-1}
\end{align}
Consistency requires following relations (and their conjugate) to
be fulfilled
\begin{align}
 & \frac{\check{b}_{\tau+1}}{(\tau+1)(\tau+s-1)}\Pi^{+}-\check{c}_{\tau}\Pi^{+}=0,\label{rec_6}\\
 & (\varsigma-s+1)(\varsigma+s+2)\check{c}_{\tau}\Pi^{+}-\frac{(\tau-s)}{(\tau+1)}\check{b}_{\tau}\Pi^{+}+\frac{(\tau-s+2)}{(\tau+1)}\check{b}_{\tau+1}\Pi^{+}-(\varsigma-s+3)(\varsigma+s)\check{c}_{\tau}\Pi^{+}=0,\label{rec_7}\\
 & 2\check{b}_{\tau}+N(\tau+1)(\tau+1+s)\check{c}_{\tau}\Pi^{+}-(N+2)(\tau-1)(\tau-1-s)\check{\bar{c}}_{\tau}\Pi^{-}-\nonumber \\
 & -(N+2)\tau(\tau+s-2)\check{c}_{\tau-1}\Pi^{+}+N\tau(\tau-s+2)\check{\bar{c}}_{\tau+1}\Pi^{-}=0,\label{rec_8}\\
 & (\varsigma+1)(\varsigma+2)(\varsigma-s+1)(\varsigma+s+2)\check{c}_{\tau}\Pi^{+}-\frac{(\tau-s+2)}{(\tau+1)}(\mu-(s-2)(s-1))\check{b}_{\tau+1}\Pi^{+}-\nonumber \\
 & -\tau(\tau-s)\check{b}_{\tau}+\tau(\tau-s+2)\check{b}_{\tau+1}-(\varsigma-s+3)(\varsigma+s)((\varsigma+1)(\varsigma+2)+\mu-(s-2)(s-1))\check{c}_{\tau}\Pi^{+}=0.\label{rec_9}
\end{align}
\eqref{rec_6} determines $\check{c}_{\tau}$ in terms of $\check{b}_{\tau}$.
Substituting this into \eqref{rec_7} one obtains an answer for $\check{b}_{\tau}\Pi^{+}$.
Combining this with a conjugate expession one can write

\begin{equation}
\check{b}_{\tau}=\frac{\check{b}}{(|\tau|+s-1)(|\tau|+s)}.
\end{equation}
Then from \eqref{rec_6} one has
\begin{equation}
\check{c}_{\tau}=\frac{\check{b}}{(\tau+1)(\tau+s-1)(\tau+s+1)(\tau+s)},
\end{equation}
and by conjugacy
\begin{equation}
\check{\bar{c}}_{\tau}=\frac{\check{b}}{(-\tau+1)(-\tau+s-1)(-\tau+s+1)(-\tau+s)}.
\end{equation}
One can always set an arbitrary constant $\check{b}$ to one $\check{b}=1$
by overall rescaling of $T$-module. Now \eqref{rec_8} holds identically,
while \eqref{rec_9} fixes the value of constant $\mu$ to be
\begin{equation}
\mu=s(s+1).
\end{equation}

Finally, one can formulate an unfolded system for the double-traceless
spin-$s$ Fronsdal current obeying \eqref{conserv_law}. This comprises
two modules $J(Y|p|x)$ and $T(Y|p,q|x)$ constrained by \eqref{s_t_limit}
and \eqref{s_t_limit_T} respectively and obeying unfolded equations
\begin{align}
 & DJ+ie\partial\bar{\partial}J+iey\bar{y}\frac{1}{(N+1)(N+2)(\bar{N}+1)(\bar{N}+2)}\biggl(-(\varsigma+1)(\varsigma+2)(\varsigma-s+1)(\varsigma+s+2)J+\nonumber \\
 & (\varsigma-s+1)(\varsigma+s+2)\frac{\partial}{\partial p}J+\frac{1}{(|\tau|+s)(|\tau|+s-1)}T\biggr)+\nonumber \\
 & +ey\bar{\partial}\frac{1}{(N+1)(N+2)}\left(\tau(\tau-s)J-\frac{(\tau-s)}{(\tau+1)}\Pi^{+}\frac{\partial}{\partial p}J-\frac{\Pi^{+}}{(\tau+1)(\tau+s+1)(\tau+s)(\tau+s-1)}T\right)+\nonumber \\
 & +e\bar{y}\partial\frac{1}{(\bar{N}+1)(\bar{N}+2)}\left(\tau(\tau+s)J-\frac{(\tau+s)}{(\tau-1)}\Pi^{-}\frac{\partial}{\partial p}J-\frac{\Pi^{-}}{(\tau-1)(\tau-s-1)(\tau-s)(\tau-s+1)}T\right)=0.\label{J_final}
\end{align}
\begin{align}
 & DT+ie\partial\bar{\partial}T+iey\bar{y}\frac{1}{(N+1)(N+2)(\bar{N}+1)(\bar{N}+2)}\biggl(-(\varsigma+1)(\varsigma+2)(\varsigma-s+\nu+3)(\varsigma+s-\nu)+\nonumber \\
 & +(\varsigma-s+\nu+3)(\varsigma+s-\nu)(\frac{\partial}{\partial p}+(\nu+2)(2s-\nu-1))+\frac{1}{(s-\nu+|\tau|-2)}\frac{\partial}{\partial q}\biggr)T+\nonumber \\
 & +ey\bar{\partial}\frac{1}{(N+1)(N+2)}\biggl(\tau(\tau-s+\nu+2)-\frac{(\tau-s+\nu+2)(\nu+2)(2s-\nu-1)}{(\tau+1)}\Pi^{+}-\nonumber \\
 & -\frac{(\tau-s+\nu+2)}{(\tau+1)}\frac{\partial}{\partial p}\Pi^{+}-\frac{1}{(\tau+1)(\tau+s-\nu-2)(\tau+s-\nu-1)}\Pi^{+}\frac{\partial}{\partial q}\biggr)T+\nonumber \\
 & +e\bar{y}\partial\frac{1}{(\bar{N}+1)(\bar{N}+2)}\biggl(\tau(\tau+s-\nu-2)-\frac{(\tau+s-\nu-2)(\nu+2)(2s-\nu-1)}{(\tau-1)}\Pi^{-}-\nonumber \\
 & -\frac{(\tau+s-\nu-2)}{(\tau-1)}\frac{\partial}{\partial p}\Pi^{-}+\frac{1}{(\tau-1)(\tau-s+\nu+2)(\tau-s+\nu+1)}\Pi^{-}\frac{\partial}{\partial q}\biggr)T=0.\label{T_final}
\end{align}

\section{Off-shell HS fields unfolded\label{SEC_OFF_SHELL_SPIN_S}}

We proceed to the final part of the main problem: in this Section
we couple unfolded Fronsdal currents \eqref{J_final}-\eqref{T_final}
to the Central On-Mass-Shell Theorem \eqref{OnShTh_1}-\eqref{OnShTh_2}
that encodes Fronsdal equations. This way in Subsection \ref{SUBSEC_COMST+CURRENT}
we arrive at unfolded system for the off-shell HS fields in $AdS_{4}$.
Then in Subsection \ref{SUBSEC_HS_QUANT} we use this system to compute
propagators of Fronsdal fields in the de Donder gauge.

\subsection{\label{SUBSEC_COMST+CURRENT}Coupling of HS currents to the Fronsdal
system}

In order to write down an Ansatz for coupling currents $J$ and $T$
to unfolded spin-$s$ Fronsdal equations \eqref{OnShTh_1}-\eqref{OnShTh_2},
one compares restrictions on $\varsigma$- and $\tau$-dependence
for $J$ and $T$ in \eqref{s_t_limit}, \eqref{s_t_limit_T} with
that for $\omega$ and $C$ in \eqref{w_C_restrictions}. Then one
can write
\begin{align}
D\omega+ey\bar{\partial}\omega+e\bar{y}\partial\omega= & \dfrac{i}{4}\eta\bar{E}\bar{\partial}\bar{\partial}C|_{N=0}+\dfrac{i}{4}\bar{\eta}E\partial\partial C|_{\bar{N}=0}+\nonumber \\
 & +\dfrac{i}{4}\bar{E}\bar{\partial}\bar{\partial}f_{N,\bar{N}}J+\dfrac{i}{4}E\partial\partial\bar{f}_{N,\bar{N}}J+iEyyg_{N,\bar{N}}T+i\bar{E}\bar{y}\bar{y}\bar{g}_{N,\bar{N}}T,\label{OnShTh_J_1}\\
DC+ie\partial\bar{\partial}C-iey\bar{y}C= & -ie\partial\bar{\partial}h_{N,\bar{N}}J-iey\bar{y}\frac{k_{N,\bar{N}}}{(N+1)(N+2)(\bar{N}+1)(\bar{N}+2)}J-\nonumber \\
 & -ey\bar{\partial}\frac{\ell_{N,\bar{N}}}{(N+1)(N+2)}J-e\bar{y}\partial\frac{\bar{\ell}_{N,\bar{N}}}{(\bar{N}+1)(\bar{N}+2)}J.\label{OnShTh_J_2}
\end{align}
Here one does not include $p$-dependent components of $J(Y|p|x)$
and $p$- or $q$-dependent components of $T(Y|p,q|x)$ to the r.h.s.
of the Ansatz, because those are identified with descendants containing
d'Alembertians or divergences of the primary sources, so they have
too high order in derivatives. Similar reasoning about the power of
derivatives excludes terms like $E\partial\partial T$ from \eqref{OnShTh_J_1}.
Such an identification, however, is not absolute and, generally speaking,
depends on the freedom in the choice of coefficients before higher-order
descendants in unfolded equations, mentioned above. An example of
a different non-diagonal coupling, including $p$-dependent components
of $J$, is presented in Appendix B.

We start with the study of consistency condition for \eqref{OnShTh_J_2}.
As follows from \eqref{w_C_restrictions}, for components of $C$,
$\tau$ takes only two values: either $s$ or $-s$. We consider $\tau=s$
case, then restoring $\tau=-s$ sector from reality. With $\tau$
fixed, $N$ and $\bar{N}$ are not independent, so $h_{N,\bar{N}}$,
$k_{N,\bar{N}}$, $\ell_{N,\bar{N}}$ and $\bar{\ell}_{N,\bar{N}}$
in \eqref{OnShTh_J_2} depend, in fact, only on one variable. It is
convenient to consider them as functions of $\varsigma$. In $\tau=s$
sector, $\bar{\ell}_{\varsigma}$ vanishes, because there is no $J$
with $\tau=s+1$, while only such $J$ could contribute under the
action of $e\bar{y}\partial$. Considering terms with $\frac{\partial}{\partial p}$
in the consistency conditions, one sees that $h_{\varsigma}$ has
to vanish as well, because through \eqref{J_final} it generates contributions
that have no counterparts to cancel with, like
\[
-\frac{1}{2}Ey\partial\frac{h_{\varsigma+1}}{(\varsigma+s+1)}\frac{\partial}{\partial p}J
\]
and others. Then one is left with a following list of consistency
conditions
\begin{align}
 & \ell_{\varsigma}-\ell_{\varsigma-1}=0,\label{rec17}\\
 & (\varsigma-s+1)(\varsigma+s+2)\ell_{\varsigma+1}-\frac{1}{s}k_{\varsigma}=0,\label{rec18}\\
 & \frac{k_{\varsigma}(s-1)}{(\varsigma+s+1)(\varsigma+s+2)(\varsigma-s+2)}-\frac{\ell_{\varsigma+1}(\varsigma+2)(\varsigma+1)(\varsigma-s+1)}{(\varsigma+s+1)(\varsigma-s+2)}+(\varsigma-s+1)\ell_{\varsigma}=0,\label{rec19}\\
 & \frac{2s^{2}\ell_{\varsigma}}{(\varsigma+s)}+\frac{k_{\varsigma-1}}{(\varsigma+s)}-\frac{k_{\varsigma}}{(\varsigma+s+2)}=0,\label{rec20}\\
 & \frac{k_{\varsigma-1}}{(\varsigma-s)}-\frac{k_{\varsigma}}{(\varsigma-s+2)}-\frac{2s^{2}\ell_{\varsigma}}{(\varsigma-s+2)}=0.\label{rec21}
\end{align}
From \eqref{rec17} it follows that
\begin{equation}
\ell_{\varsigma}=\ell
\end{equation}
is constant. Then \eqref{rec18} yields
\begin{equation}
k_{\varsigma}=s(\varsigma-s+1)(\varsigma+s+2)\ell\label{m_part}
\end{equation}
and rest of equations \eqref{rec19}-\eqref{rec21} hold identically.

Now one has to determine coupling coefficients $f$ and $g$ in \eqref{OnShTh_J_1}.
First, one notes that due to \eqref{w_C_restrictions} $f$ and $g$
are functions of $\tau$ only. Then, analyzing consistency relations
for \eqref{OnShTh_J_1} that involve $\frac{\partial}{\partial p}$,
one sees that all four terms with $\frac{\partial}{\partial p}$ are
different and do not cancel, so the only way out is to put coefficients
before them to zero, which amounts to
\begin{equation}
f_{\tau<0}=0,\quad\bar{f}_{\tau>0}=0,\quad g_{\tau<0}=0,\quad\bar{g}_{\tau>0}=0.\label{f_g_chiral}
\end{equation}
Then consistency conditions read
\begin{align}
 & \frac{N\tau(\tau+s-2)}{(\bar{N}+1)(\bar{N}+2)}g_{\tau-1}+\frac{\bar{N}\tau(\tau-s+2)}{(N+1)(N+2)}\bar{g}_{\tau+1}-(N+3)g_{\tau}-(\bar{N}+3)\bar{g}_{\tau}=0,\label{rec22}\\
 & \frac{f_{\tau}}{(N+1)(N+2)(\bar{N}+1)(|\tau|+s)(|\tau|+s-1)}+Ng_{\tau}=0,\label{rec23}\\
 & \frac{\tau(\tau+s)}{(\bar{N}+1)}f_{\tau-1}-(\bar{N}-1)f_{\tau}-\frac{\bar{\eta}\ell}{(N+1)}|_{\bar{N}=1}=0,\label{rec24}
\end{align}
plus conjugate ones. From \eqref{rec22} one finds, taking into account
\eqref{w_C_restrictions} and \eqref{f_g_chiral},
\begin{align}
 & g_{\tau}=g\frac{(s+\tau-2)!^{2}(s-\tau-2)!(s-\tau-1)!\tau!}{(s+\tau+1)!}\Pi^{+},\\
 & \bar{g}_{\tau}=g\frac{(s+|\tau|-2)!^{2}(s-|\tau|-2)!(s-|\tau|-1)!|\tau|!}{(s+|\tau|+1)!}\Pi^{-}
\end{align}
with $g$ being arbitrary coupling constant. Now \eqref{rec23} yields
\begin{equation}
f_{\tau}=-g(s-\tau-2)!(s-\tau)!(s+\tau)!\tau!\Pi^{+}.
\end{equation}
Finally, \eqref{rec24} expresses $\ell$ in terms of the coupling
constant $g$ and the phase factor $\eta$ as
\begin{equation}
\ell=-2g\eta(2s-1)!s!.
\end{equation}

So a consistent coupling of $J$ and $T$ to $\omega$ and $C$ is
\begin{align}
D\omega+ey\bar{\partial}\omega+e\partial\bar{y}\omega= & \biggl(\dfrac{i}{4}\eta\bar{E}\bar{\partial}\bar{\partial}C|_{N=0}+\dfrac{g}{4}\bar{E}\bar{\partial}\bar{\partial}(s-\tau-2)!(s-\tau)!(s+\tau)!\tau!\Pi^{+}J-\nonumber \\
 & -\frac{g}{4}Eyy\frac{(s+\tau-2)!^{2}(s-\tau-2)!(s-\tau-1)!\tau!}{(s+\tau+1)!}\Pi^{+}T+h.c.\biggr)\label{omega_OffShell}\\
DC+ie\partial\bar{\partial}C-iey\bar{y}C= & -g\left(iey\bar{y}\frac{\eta(2s)!s!}{(\varsigma+s+1)(\varsigma-s+2)}\Pi^{+}J+ey\bar{\partial}\frac{2\eta(2s-1)!s!}{(\varsigma+s)(\varsigma+s+1)}\Pi^{+}J+h.c.\right)\label{C_OffShell}
\end{align}
This completes unfolding of the off-shell HS fields. Equations \eqref{omega_OffShell}-\eqref{C_OffShell}
and \eqref{J_final}-\eqref{T_final} provide an unfolded formulation
of a massless spin-$s$ field coupled to the external Fronsdal current.
Interpreted differently, they form an off-shell completion of the
Central On-Mass-Shell Theorem \eqref{OnShTh_1}-\eqref{OnShTh_2}:
the only primary field is $\omega_{s-1,s-1}(Y|x)$, which encodes
unconstrained double-traceless spin-$s$ Fronsdal field and generates
Fronsdal gauge symmetry through \eqref{unf_gauge_transf}, while other
components of $\omega$ as well as the whole modules $C$, $J$ and
$T$ form a complete basis of its descendants. A reverse on-shell
reduction of \eqref{omega_OffShell}-\eqref{C_OffShell} to the unfolded
Fronsdal equations \eqref{OnShTh_1}-\eqref{OnShTh_2} is trivially
achieved by sending the coupling constant to zero $g=0.$

Translating into the language of representation theory, we did the
following: we took $SO(3,2)$-module $D(s+1,s)$ (accompanied by an
appropriate gauge module in the 1-form sector), which corresponds
to an $AdS_{4}$ massless spin-$s$ field, and \textquotedbl glued\textquotedbl{}
it to the Fronsdal current module corresponding to $J$ and $T$,
in order to get an indecomposable off-shell spin-$s$ Verma module.
This resulting module has more complicated structure in comparison
with the scalar off-shell module constructed in Section \ref{SEC_SPIN_0}
and does not allow for the same straightforward and simple analysis.
However, the most distinctive features remain the same: the representation
is infinitely indecomposable; the module represents an infinite sequence
of nested submodules; quotient by some submodule leads to an on-shell
system with (in general higher-derivative) e.o.m.; reduction to the
initial on-shell Fronsdal system amounts to the quotient by the maximal
submodule dual to entire $J$ and $T$.

Now we will demonstrate how one can use the off-shell system we built
in order to quantize the theory. Namely, we will evaluate HS propagators
in the de Donder gauge from \eqref{omega_OffShell}, similarly to
what we did in the scalar field example in Subsection \ref{SUBSEC_SCAL_QUANT}.

\subsection{Quantization: massless HS propagators in the de Donder gauge\label{SUBSEC_HS_QUANT}}

Having in hand unfolded equations for off-shell Fronsdal fields and
conjugate operators extracting wave equations therefrom, one can compute
correlation functions.

To this end one first fixes the gauge. Making use of the unfolded
gauge transformation \eqref{unf_gauge_transf}, which for \eqref{omega_OffShell}
takes the form
\begin{equation}
\delta\omega=D\varepsilon+(ey\bar{\partial})\varepsilon+(e\bar{y}\partial)\varepsilon
\end{equation}
and in terms of constituent 0-forms \eqref{w_decompos} reads as
\begin{equation}
\delta\phi=\frac{1}{N\bar{N}}(y^{\alpha}\bar{y}^{\dot{\alpha}}D_{\alpha\dot{\alpha}})\varepsilon,\quad\delta\widetilde{\phi}=\frac{1}{(N+2)(\bar{N}+2)}(\partial^{\alpha}\bar{\partial}^{\dot{\alpha}}D_{\alpha\dot{\alpha}})\varepsilon,
\end{equation}
one can impose generalized de Donder gauge
\begin{equation}
(\partial^{\alpha}\bar{\partial}^{\dot{\alpha}}D_{\alpha\dot{\alpha}})\phi-(y^{\alpha}\bar{y}^{\dot{\alpha}}D_{\alpha\dot{\alpha}})\widetilde{\phi}=0.\label{de_Donder_gauge}
\end{equation}
Then, applying conjugate operators \eqref{D*_phi}, \eqref{D*_phi_tilde}
to \eqref{omega_OffShell} and accounting for \eqref{de_Donder_gauge},
one has in the primary $\tau=0$ sector
\begin{equation}
\left(\square-\lambda^{2}(s^{2}-2s-2)\right)\phi_{\alpha(s),\dot{\alpha}(s)}=\frac{g}{2}(s-1)!^{2}s!J_{\alpha(s),\dot{\alpha}(s)},
\end{equation}
\begin{equation}
\left(\square-\lambda^{2}(s^{2}+2s-2)\right)\widetilde{\phi}_{\alpha(s-2),\dot{\alpha}(s-2)}=\frac{g}{2s}(s-2)!^{3}T_{\alpha(s-2),\dot{\alpha}(s-2)}.
\end{equation}
Replacing $\phi_{\alpha(s),\dot{\alpha}(s)}\rightarrow\dfrac{\delta W}{\delta J^{\alpha(s),\dot{\alpha}(s)}}$,
$\widetilde{\phi}_{\alpha(s-2),\dot{\alpha}(s-2)}\rightarrow\dfrac{\delta W}{\delta T^{\alpha(s-2),\dot{\alpha}(s-2)}}$,
one restores propagators for traceless and trace parts of the Fronsdal
spin-$s$ field in the de Donder gauge
\begin{equation}
\bigl\langle\phi_{\alpha(s),\dot{\alpha}(s)}(x)\phi_{\beta(s),\dot{\beta}(s)}(x')\bigr\rangle=\frac{g}{2}(s-1)!^{2}s!(\epsilon_{\alpha\beta})^{s}(\epsilon_{\dot{\alpha}\dot{\beta}})^{s}G_{\Delta_{s}}(x,x'),
\end{equation}
\begin{equation}
\bigl\langle\widetilde{\phi}_{\alpha(s-2),\dot{\alpha}(s-2)}(x)\widetilde{\phi}_{\beta(s-2),\dot{\beta}(s-2)}(x')\bigr\rangle=\frac{g}{2s}(s-2)!^{3}(\epsilon_{\alpha\beta})^{s-2}(\epsilon_{\dot{\alpha}\dot{\beta}})^{s-2}G_{\widetilde{\Delta}_{s}}(x,x'),
\end{equation}
where $G_{\Delta}(x,x')$ is the scalar Green function \eqref{Green_function}
with conformal weights determined from 
\begin{equation}
\Delta_{s}(\Delta_{s}-3)-s=s^{2}-2s-2,\quad\widetilde{\Delta}_{s}(\widetilde{\Delta}_{s}-3)-(s-2)=s^{2}+2s-2.
\end{equation}

\section{Conclusion\label{SEC_CONCLUSIONS}}

In the paper we have constructed an off-shell completion for unfolded
on-shell system of free massless bosonic fields in $AdS_{4}$. This
has been done by coupling them to the external HS currents, which
from the standpoint of unfolded approach represent some special set
of descendants of primary Fronsdal fields. We have constructed an
appropriate unfolded system, which describes these HS currents. The
process of construction goes as follows: first, one determines an
appropriate unfolded module, which should include all possible descendants
(excluding those vanishing due to differential constraints) of a primary
field; then one writes down a suitable Ansatz for unfolded equations;
finally, one fixes coefficients in the Ansatz, partially by solving
for consistency constraints and partially based on convenience reasons
(this reflects a large freedom in choice of basis in the descendant
space). It turns out that this basis choice may seriously affect a
further analysis: although any option leads to a consistent unfolded
system of HS currents, thus solving the problem, not any such system
can be diagonally coupled to unfolded Fronsdal equations. By this
we mean that in general some linear combination of unfolded currents
appears in the r.h.s. of the Fronsdal equation, and only in certain
cases, achieved by a particular smart choice of coefficients, there
will be a single unfolded current, that can be therefore unambiguously
identified with the Fronsdal source.

The possibility of such identification is of particular importance
if one is about to study the quantum aspects of the off-shell unfolded
system. As we have demonstrated, classical unfolded off-shell system
can be reformulated as the set of Schwinger\textendash Dyson equations
that determine a partition function of the underlying quantum theory.
To this end one should treat primary unfolded fields and primary unfolded
currents as conjugated variables upon action on a partition function.
This way we have managed to restore $AdS_{4}$-propagators for Fronsdal
fields in the de Donder gauge. Although we have not presented here
a rigorous comprehensive prescription for quantization of a general
unfolded field theory, which should be a topic for a separate thorough
study, our analysis demonstrates at least the principal possibility
of extracting quantum answers from an off-shell unfolded system.

This paves the way for exploring quantum features of nonlinear higher-spin
gravity. Since the on-shell HS system we started with arises from
a linearization of Vasiliev equations, its off-shell extension we
have constructed should represent a linear limit of a would-be off-shell
completion of Vasiliev theory. Constructing this off-shell completed
Vasiliev theory together with formulating general quantization prescription
for unfolded systems will allow one to start systematic investigation
of quantum HS gravity and, among other, hopefully will help to proceed
in the most urgent problem of spacetime- and spin-locality of HS interactions.

From the point of view of representation theory, the off-shell system
we have constructed corresponds to an indecomposable representation
of $SO(3,2)$, which leads to the initial on-shell system after quotienting
by a maximal submodule, which is the external current module. This
off-shell representation is in fact infinitely indecomposable, being
an infinite sequence of successively nested submodules. It would be
interesting to explore the structure of the off-shell module in more
detail, in order to clarify the representation-theory picture of our
construction.

As a byproduct of our analysis, we have also discovered a simple way
of extracting wave equations for component fields from an unfolded
system by means of \textquotedbl conjugate operators\textquotedbl{}
$D^{*}$ \eqref{D*_phi}, \eqref{D*_phi_tilde}, \eqref{D*_C_m},
\eqref{D*_FP} and have found an interesting reduction of the unfolded
system for HS currents, which turns it to the unfolded Fierz-Pauli
system \eqref{Fierz-Pauli} for massive HS fields.

\section*{Acknowledgments}

The author is grateful to M.A. Grigoriev, E. Joung, E.D. Skvortsov
and M.A. Vasiliev for valuable discussions and useful comments, and
to the Referee for important remarks on the structure of the off-shell
modules. The research was supported by the Russian Science Foundation
grant 18-12-00507.

\section*{Appendix A\label{Appendix A}. Notations and conventions}

Here we list conventions used in calculations and collect notations
introduced in the paper.

We deal with $4d$ anti-de Sitter space $AdS_{4}$ with a negative
cosmological constant $\Lambda$ and an inverse radius $\lambda$
such that $\Lambda=-3\lambda^{2}$. Tensor indices referring to this
space are denoted by underlined lowercase Latin letters $\underline{a},\underline{b},\underline{c},...=\left\{ 0,1,2,3\right\} $
and are transforming by global $AdS$ symmetry group $SO(3,2)$. $g_{\underline{ab}}$
is $AdS$ metric, $D_{\underline{a}}$ is $AdS$-covariant derivative
and $\square=D_{\underline{a}}D^{\underline{a}}$ is covariant d'Alembertian.

To work with unfolded modules we introduce a flat fiber space which
is linked to the base $AdS_{4}$-manifold through a vierbein field
$e_{\underline{a}}^{\alpha\dot{\beta}}\left(x\right)$, where $\alpha$
and $\dot{\beta}$, taking two values $\left\{ 1,2\right\} $, correspond
to two spinor representations of the fiber Lorentz algebra $so(3,1)\approx sp(2,\mathbb{C})$.
Raising and lowering of spinor indices are carried out by means of
antisymmetric spinor metric
\[
\epsilon_{\alpha\beta}=\epsilon_{\dot{\alpha}\dot{\beta}}=\left(\begin{array}{cc}
0 & 1\\
-1 & 0
\end{array}\right),\quad\epsilon^{\alpha\beta}=\epsilon^{\dot{\alpha}\dot{\beta}}=\left(\begin{array}{cc}
0 & 1\\
-1 & 0
\end{array}\right),\eqno(\text{\textcyr{\CYRA}}.1)
\]
according to conventions
\[
v_{\alpha}=\epsilon_{\beta\alpha}v^{\beta},\quad v^{\alpha}=\epsilon^{\alpha\beta}v_{\beta},\quad\bar{v}_{\dot{\alpha}}=\epsilon_{\dot{\beta}\dot{\alpha}}\bar{v}^{\dot{\beta}},\quad\bar{v}^{\dot{\alpha}}=\epsilon^{\dot{\alpha}\dot{\beta}}\bar{v}_{\dot{\beta}}.\eqno(\text{\textcyr{\CYRA}}.2)
\]
For vierbein we fix the following normalization
\[
e_{\underline{a}}^{\alpha\dot{\alpha}}e_{\underline{b}}^{\beta\dot{\beta}}\epsilon_{\alpha\beta}\epsilon_{\dot{\alpha}\dot{\beta}}=-\frac{1}{2}g_{\underline{ab}},\quad e_{\underline{a}}^{\alpha\dot{\alpha}}e_{\underline{b}}^{\beta\dot{\beta}}g^{\underline{ab}}=-\frac{1}{2}\epsilon^{\alpha\beta}\epsilon^{\dot{\alpha}\dot{\beta}}.\eqno(\text{\textcyr{\CYRA}}.3)
\]
We define a dimensionless covariant derivative $D_{\alpha\dot{\beta}}$
as
\[
D^{\alpha\dot{\beta}}=-\frac{2}{\lambda}e_{\underline{a}}^{\alpha\dot{\beta}}D^{\underline{a}}\quad\Longrightarrow\quad D=\mathrm{d}x^{\underline{a}}e_{\underline{a}}^{\alpha\dot{\beta}}D_{\alpha\dot{\beta}},\quad\square=-\frac{\lambda^{2}}{2}D_{\alpha\dot{\beta}}D^{\alpha\dot{\beta}}.\eqno(\text{\textcyr{\CYRA}}.4)
\]
We make use of condensed notations for higher-rank symmetric tensors
and multispinors, so
\[
T_{\underline{a}(n)}=T_{\underline{a_{1}}\underline{a_{2}}...\underline{a_{n}}},\quad T_{\alpha(n),\dot{\beta}(m)}=T_{\alpha_{1}\alpha_{2}...\alpha_{n},\dot{\beta}_{1}\dot{\beta}_{2}...\dot{\beta}_{m}}.\eqno(\text{\textcyr{\CYRA}}.5)
\]
To deal with unfolded fields we introduce a pair of auxiliary $sp(2,\mathbb{C})$-spinors
$Y=\left\{ y^{\alpha},\bar{y}^{\dot{\alpha}}\right\} $ which are
commutative variables, so 
\[
y^{\alpha}y^{\beta}\epsilon_{\alpha\beta}=0,\quad\bar{y}^{\dot{\alpha}}\bar{y}^{\dot{\beta}}\epsilon_{\dot{\alpha}\dot{\beta}}=0,\eqno(\text{\textcyr{\CYRA}}.6)
\]
and corresponding derivatives
\[
\partial_{\alpha}y^{\beta}=\delta_{\alpha}\text{}^{\beta},\quad\bar{\partial}_{\dot{\alpha}}\bar{y}^{\dot{\beta}}=\delta_{\dot{\alpha}}\text{}^{\dot{\beta}}.\eqno(\text{\textcyr{\CYRA}}.7)
\]
From \eqref{D_definition} and \eqref{ads1}-\eqref{ads2} one finds
for an $AdS$-derivatives commutator acting on some unfolded module
$F(Y|x)$
\[
[D_{\alpha\dot{\alpha}},D_{\beta\dot{\beta}}]F=-\epsilon_{\alpha\beta}(\bar{y}_{\dot{\alpha}}\bar{\partial}_{\dot{\beta}}+\bar{y}_{\dot{\beta}}\bar{\partial}_{\dot{\alpha}})F-\epsilon_{\dot{\alpha}\dot{\beta}}(y_{\alpha}\partial_{\beta}+y_{\beta}\partial_{\alpha})F.\eqno(\text{\textcyr{\CYRA}}.8)
\]
$Y$-counting operators $N$ and $\bar{N}$ and their linear combinations
$\varsigma$ and $\tau$ are
\[
N=y^{\alpha}\partial_{\alpha},\quad\bar{N}=\bar{y}^{\dot{\alpha}}\bar{\partial}_{\dot{\alpha}},\quad\varsigma=\frac{1}{2}(N+\bar{N}),\quad\tau=\frac{1}{2}(N-\bar{N}).\eqno(\text{\textcyr{\CYRA}}.9)
\]
$\Pi^{+}$ and $\Pi^{-}$ are projectors on non-negative and non-positive
$\tau$ components of the unfolded module
\[
\Pi^{+}F_{N,\bar{N}}(Y)=\begin{cases}
F_{N,\bar{N}}(Y), & \tau\geq0\\
0, & \tau<0
\end{cases};\qquad\Pi^{-}F_{N,\bar{N}}(Y)=\begin{cases}
F_{N,\bar{N}}(Y), & \tau\leq0\\
0, & \tau>0
\end{cases}.\eqno(\text{\textcyr{\CYRA}}.10)
\]
To simplify notations we omit contracted indices between vierbein
1-form $e^{\alpha\dot{\beta}}=\mathrm{d}x^{\underline{a}}e_{\underline{a}}^{\alpha\dot{\beta}}$
and $\left\{ y,\bar{y},\partial,\bar{\partial}\right\} $ in unfolded
equations, so that
\[
ey\bar{y}=e^{\alpha\dot{\beta}}y_{\alpha}\bar{y}_{\dot{\beta}},\quad e\partial\bar{\partial}=e^{\alpha\dot{\beta}}\partial_{\alpha}\bar{\partial}_{\dot{\beta}},\quad ey\bar{\partial}=e^{\alpha\dot{\beta}}y_{\alpha}\bar{\partial}_{\dot{\beta}},\quad e\bar{y}\partial=e^{\alpha\dot{\beta}}\bar{y}_{\dot{\beta}}\partial_{\alpha},\eqno(\text{\textcyr{\CYRA}}.11)
\]
and for basis 2-forms
\[
E^{\alpha\beta}=e^{\alpha}\text{}_{\dot{\gamma}}e^{\beta\dot{\gamma}},\quad\bar{E}^{\dot{\alpha}\dot{\beta}}=e_{\gamma}\text{}^{\dot{\alpha}}e^{\gamma\dot{\beta}}\eqno(\text{\textcyr{\CYRA}}.12)
\]
the same \textquotedbl top-down\textquotedbl{} rules for contractions
with $\left\{ y,\bar{y},\partial,\bar{\partial}\right\} $ hold.

\section*{Appendix B.\label{Appendix B} Non-diagonal spin-s off-shell extension}

One can study \eqref{rec3}-\eqref{rec1} without fixing $a_{N,\bar{N}}=1$
as in \eqref{a_1}. Then after some tedious algebra one arrives at
following consistency relations
\[
a_{N+1,\bar{N}+1}b_{N,\bar{N}}=\frac{(\varsigma+1)(\varsigma+2)(\varsigma-s+1)(\varsigma+s+2)}{(N+1)(N+2)(\bar{N}+1)(\bar{N}+2)},\eqno(\text{B}.1)
\]
\[
c_{N,\bar{N}+1}\bar{c}_{N+1,\bar{N}}=\frac{(\tau-\frac{1}{2})(\tau+\frac{1}{2})(\tau-s-\frac{1}{2})(\tau+s+\frac{1}{2})}{(N+1)(N+2)(\bar{N}+1)(\bar{N}+2)},\eqno(\text{B}.2)
\]
and finds that $a_{N,\bar{N}}$, $b_{N,\bar{N}}$, $c_{N,\bar{N}}$
and $\bar{c}_{N,\bar{N}}$ can be presented as
\[
a_{N,\bar{N}}=a_{\varsigma}\alpha_{N}\bar{\alpha}_{\bar{N}},\quad b_{N,\bar{N}}=b_{\varsigma}\beta_{N}\bar{\beta}_{\bar{N}},\quad\bar{c}_{N,\bar{N}}=\bar{c}_{\tau}\alpha_{N}\bar{\beta}_{\bar{N}},\quad c_{N,\bar{N}}=c_{\tau}\bar{\alpha}_{\bar{N}}\beta_{N},\eqno(\text{B}.3)
\]
with the following constraints
\[
(N+2)\alpha_{N+1}\beta_{N}=N\alpha_{N}\beta_{N-1},\quad(\bar{N}+2)\bar{\alpha}_{\bar{N}+1}\bar{\beta}_{\bar{N}}=\bar{N}\bar{\alpha}_{\bar{N}}\bar{\beta}_{\bar{N}-1},\eqno(\text{B}.4)
\]
\[
a_{\varsigma+1}b_{\varsigma}=(\varsigma+1)(\varsigma+2)(\varsigma-s+1)(\varsigma+s+2),\eqno(\text{B}.5)
\]
\[
c_{\tau-\frac{1}{2}}\bar{c}_{\tau+\frac{1}{2}}=(\tau-\frac{1}{2})(\tau+\frac{1}{2})(\tau-s-\frac{1}{2})(\tau+s+\frac{1}{2}).\eqno(\text{B}.6)
\]
One can choose a partial solution to this system which provides a
more simpler-looking form (with more symmetric coefficients) for the
equations on $J$ than \eqref{massless_J}. Then instead of \eqref{J_conserved_eq}
one arrives at
\begin{align*}
 & DJ+ie\partial\bar{\partial}\frac{(\varsigma+s+1)}{(N+1)(\bar{N}+1)}\left(\varsigma+\frac{\partial}{\partial p}+1\right)J-iey\bar{y}\frac{(\varsigma-s+1)}{(N+1)(\bar{N}+1)}\left(\varsigma-\frac{\partial}{\partial p}+1\right)J+\\
 & +e\bar{y}\partial\frac{(\tau+s)}{(N+1)(\bar{N}+1)}\left(\tau+\frac{\partial}{\partial p}\right)J+ey\bar{\partial}\frac{(\tau-s)}{(N+1)(\bar{N}+1)}\left(\tau-\frac{\partial}{\partial p}\right)J=0. &  &  &  & (\text{B}.7)
\end{align*}
However, coupling of this current to the Central On-Mass-Shell-Theorem
takes the form
\begin{align*}
 & (D+ey\bar{\partial}+e\partial\bar{y})\omega=\dfrac{i\bar{\eta}}{4}E\partial\partial\left(C-f(2s)!J\right)|_{\bar{N}=0}+\dfrac{ig}{4}E\partial\partial\frac{(s-2\tau-\frac{\partial}{\partial p})}{(s+\tau+1)(s-\tau+1)}J+h.c., & (\text{B}.8)\\
 & (D+ie\partial\bar{\partial}-iey\bar{y})C=ifey\bar{y}\frac{(\varsigma+s)!}{(\varsigma-s+2)!}(s-\frac{\partial}{\partial p}-1)\Pi^{+}J+\\
 & +ife\partial\bar{\partial}\frac{(\varsigma+s-1)!}{(\varsigma-s+1)!}(s-\frac{\partial}{\partial p}-1)\Pi^{+}J+fey\bar{\partial}\frac{(\varsigma+s-1)!}{(\varsigma-s+2)!}(s-\frac{\partial}{\partial p}-1)\Pi^{+}J+h.c. & (\text{B}.9)
\end{align*}
with arbitrary unrelated coupling constants $g$, $f$ and $\bar{f}$.
From here one sees that $J_{\alpha(s),\dot{\alpha}(s)}(x)|_{p=0}$
cannot be unambiguously identified with the primary Fronsdal current
because $J$ linear in $p$ also contributes to the r.h.s. of Fronsdal
equations. Moreover, now $C$ cannot be strictly identified with on-shell
d.o.f. of Fronsdal field because $J$ arises at the same places in
$\omega$-equations as $C$ does.Thus, although this system also provides
an off-shell unfolded formulation for spin-$s$ field, an interpretation
of different descendants is obscure, that, in particular, obstructs
the procedure of quantization. So one concludes that the basis of
descendants fixed by this choice of coefficients $a_{N,\bar{N}}$,
$b_{N,\bar{N}}$, $c_{N,\bar{N}}$ and $\bar{c}_{N,\bar{N}}$ is \textquotedbl non-diagonal\textquotedbl .

The reason behind this is just the excessive symmetry of coefficients
in (B.7), which requires $\frac{\partial}{\partial p}$ to be presented
in all terms. Because of this, already the very first equation for
$J_{\alpha(s),\dot{\alpha}(s)}(x)|_{p=0}$ involves component of $J$
linear in $p$, that obstructs local expression of $p$-dependent
components in terms of the $p$-independent ones. From the standpoint
of representation theory, it means that the module set by (B.7) is
not of a lowest-weight type. Thus the primary Fronsdal current turns
to be smeared over two $J_{\alpha(s),\dot{\alpha}(s)}$, $p$-independent
and $p$-linear, and this is indeed what one sees from (B.8). Of course,
since both \textquotedbl non-diagonal\textquotedbl{} (B.7) and \textquotedbl diagonal\textquotedbl{}
\eqref{J_conserved_eq} describe the same dynamical system (spin-$s$
conserved current), they must be related by some field redefinition,
but this would be severely non-local in terms of $Y$ and $p$.

Thus, among the large set of formally consistent unfolded systems
for conserved spin-$s$ current, only a special class corresponds
to the lowest-weight modules, which can be diagonally coupled to the
Fronsdal system. And the choice \eqref{a_1}, \eqref{vanish_a_c_c-}
fixes a particular representative from this class.


\begin{thebibliography}{10}
\bibitem{vas1}M.A. Vasiliev, \emph{Phys.Lett.B} \textbf{243} (1990)
378-382.

\bibitem{vas2}M.A. Vasiliev, \emph{Phys.Lett.B} \textbf{285} (1992)
225-234.

\bibitem{unf2}M.A. Vasiliev, \emph{Class.Quant.Grav.} \textbf{11}
(1994) 649-664.

\bibitem{cub1}S. Giombi, Xi Yin, \emph{JHEP} \textbf{1009} (2010)
115 \href{https://arxiv.org/abs/0912.3462}{[arXiv:0912.3462]}.

\bibitem{key-4}S. Giombi, Xi Yin, \emph{JHEP} \textbf{1104} (2011)
086 \href{https://arxiv.org/abs/1004.3736}{[arXiv:1004.3736]}.

\bibitem{key-5}V.E. Didenko, M.A. Vasiliev, \emph{Phys.Lett.B} \textbf{775}
(2017) 352-360 \href{https://arxiv.org/abs/1705.03440}{[arXiv:1705.03440]}.

\bibitem{key-6}E. Sezgin, E.D. Skvortsov, Y. Zhu, \emph{JHEP} \textbf{1707}
(2017) 133 \href{https://arxiv.org/abs/1705.03197}{[arXiv:1705.03197]}.

\bibitem{key-7}O.A. Gelfond, M.A. Vasiliev, \emph{Nucl.Phys.B} \textbf{931}
(2018) 383-417 \href{https://arxiv.org/abs/1706.03718}{[arXiv:1706.03718]}.

\bibitem{cub2}N.G. Misuna, \emph{Phys.Lett.B} \textbf{778} (2018)
71-78 \href{https://arxiv.org/abs/1706.04605}{[arXiv:1706.04605]}.

\bibitem{quart1}V.E. Didenko, O.A. Gelfond, A.V. Korybut, M.A. Vasiliev,
\emph{J.Phys.A} \textbf{51} (2018) 46, 465202 \href{https://arxiv.org/abs/1807.00001}{[arXiv:1807.00001]}.

\bibitem{key-10}V.E. Didenko, O.A. Gelfond, A.V. Korybut, M.A. Vasiliev,
\emph{JHEP} \textbf{12} (2019) 086 \href{https://arxiv.org/abs/1909.04876}{[arXiv:1909.04876]}.

\bibitem{quart2}V.E. Didenko, O.A. Gelfond, A.V. Korybut, M.A. Vasiliev,
\emph{JHEP} \textbf{12} (2020) 184 \href{https://arxiv.org/abs/2009.02811}{[arXiv:2009.02811]}.

\bibitem{AdSCFT1}I. R. Klebanov, A. M. Polyakov, \emph{Phys.Lett.B}
\textbf{550} (2002) 213-219 \href{https://arxiv.org/abs/hep-th/0210114}{[hep-th/0210114]}.

\bibitem{key-13}E. Sezgin, P. Sundell, \emph{Nucl.Phys.B} \textbf{644}
(2002) 303-370, \emph{Nucl.Phys.B} \textbf{660} (2003) 403-403 (erratum)
\href{https://arxiv.org/abs/hep-th/0205131}{[hep-th/0205131]}.

\bibitem{key-14}R.G. Leigh, A.C. Petkou, \emph{JHEP} \textbf{06}
(2003) 011 \href{https://arxiv.org/abs/hep-th/0304217}{[hep-th/0304217]}.

\bibitem{key-15}E. Sezgin, P. Sundell, \emph{JHEP} \textbf{07} (2005)
044 \href{https://arxiv.org/abs/hep-th/0305040}{[hep-th/0305040]}.

\bibitem{key-16}S. Giombi, S. Minwalla, S. Prakash, S.P. Trivedi,
S.R. Wadia, Xi Yin, \emph{Eur.Phys.J.C} \textbf{72} (2012) 2112 \href{https://arxiv.org/abs/1110.4386}{[arXiv:1110.4386]}.

\bibitem{AdSCFT2}O. Aharony, G. Gur-Ari, R. Yacoby, \emph{JHEP} \textbf{03}
(2012) 037 \href{https://arxiv.org/abs/1110.4382}{[arXiv:1110.4382]}.

\bibitem{ActionsCharges}M.A. Vasiliev, \emph{Int.J.Geom.Meth.Mod.Phys.}
\textbf{3} (2006) 37-80 \href{https://arxiv.org/abs/hep-th/0504090}{[hep-th/0504090]}.

\bibitem{WZ}N.G. Misuna, M.A. Vasiliev, \emph{JHEP} \textbf{05} (2014)
140 \href{https://arxiv.org/abs/1301.2230}{[arXiv:1301.2230]}.

\bibitem{act1}S.R. Das, A. Jevicki, \emph{Phys.Rev.D} \textbf{68}
(2003) 044011 \href{https://arxiv.org/abs/hep-th/0304093}{[hep-th/0304093]}.

\bibitem{act2}A. Fotopoulos, M. Tsulaia, \emph{Int.J.Mod.Phys.A}
\textbf{24} (2009) 1-60 \href{https://arxiv.org/abs/0805.1346}{[arXiv:0805.1346]}.

\bibitem{act3}A. Jevicki, K. Jin, Q. Ye, \emph{J.Phys.A} \textbf{44}
(2011) 465402 \href{https://arxiv.org/abs/1106.3983}{[arXiv:1106.3983]}.

\bibitem{act4}N. Boulanger, P. Sundell, \emph{J.Phys.A} \textbf{44}
(2011) 495402 \href{https://arxiv.org/abs/1102.2219}{[arXiv:1102.2219]}.

\bibitem{act5}N. Boulanger, N. Colombo, P. Sundell, \emph{JHEP} \textbf{10}
(2012) 043 \href{https://arxiv.org/abs/1205.3339}{[arXiv:1205.3339]}.

\bibitem{act6}N. Boulanger, E. Sezgin, P. Sundell, 4D Higher Spin
Gravity with Dynamical Two-Form as a Frobenius-Chern-Simons Gauge
Theory \href{https://arxiv.org/abs/1505.04957}{[arXiv:1505.04957]}.

\bibitem{act7}I.L. Buchbinder, K. Koutrolikos, \emph{JHEP} \textbf{12}
(2015) 106 \href{https://arxiv.org/abs/1510.06569}{[arXiv:1510.06569]}.

\bibitem{act8}C. Arias, R. Bonezzi, N. Boulanger, E. Sezgin, P. Sundell,
in \emph{Proccedings of International Workshop on Higher Spin Gauge
Theories, 4-6 November 2015, Singapore} (2017) 213-253, \href{https://arxiv.org/abs/1603.04454}{[arXiv:1603.04454]}.

\bibitem{Z1}S. Giombi, I.R. Klebanov, \emph{JHEP} \textbf{12} (2013)
068 \href{https://arxiv.org/abs/1308.2337}{[arXiv:1308.2337]}.

\bibitem{Z2}S. Giombi, I.R. Klebanov, B.R. Safdi, \emph{Phys.Rev.D}
\textbf{89} (2014) 8, 084004 \href{https://arxiv.org/abs/1401.0825}{[arXiv:1401.0825]}.

\bibitem{Z3}A. Jevicki, K. Jin, J. Yoon, \emph{Phys.Rev.D} \textbf{89}
(2014) 8, 085039 \href{https://arxiv.org/abs/1401.3318}{[arXiv:1401.3318]}.

\bibitem{Z4}S. Giombi, I.R. Klebanov, A.A. Tseytlin, \emph{Phys.Rev.D}
\textbf{90} (2014) 2, 024048 \href{https://arxiv.org/abs/1402.5396}{[arXiv:1402.5396]}.

\bibitem{Z5}M. Beccaria, A.A. Tseytlin, \emph{JHEP} \textbf{11} (2014)
114 \href{https://arxiv.org/abs/1410.3273}{[arXiv:1410.3273]}.

\bibitem{Z6}M. Beccaria, A.A. Tseytlin, \emph{J.Phys.A} \textbf{48}
(2015) 27, 275401 \href{https://arxiv.org/abs/1503.08143}{[arXiv:1503.08143]}.

\bibitem{Z7}M. Günaydin, E.D. Skvortsov, T. Tran, \emph{JHEP} \textbf{11}
(2016) 168 \href{https://arxiv.org/abs/1608.07582}{[arXiv:1608.07582]}.

\bibitem{Z8}Yi Pang, E. Sezgin, Y. Zhu, \emph{Phys.Rev.D} \textbf{95}
(2017) 2, 026008 \href{https://arxiv.org/abs/1608.07298}{[arXiv:1608.07298]}.

\bibitem{Z9}S. Giombi, I.R. Klebanov, Z.M. Tan, \emph{Universe} \textbf{4}
(2018) 1, 18 \href{https://arxiv.org/abs/1608.07611}{[arXiv:1608.07611]}.

\bibitem{Z10}E.D. Skvortsov, T. Tran, \emph{Universe} \textbf{3}
(2017) 3, 61 \href{https://arxiv.org/abs/1707.00758}{[arXiv:1707.00758]}.

\bibitem{amp0}D. Ponomarev, A.A. Tseytlin, \emph{JHEP} \textbf{05}
(2016) 184, \href{https://arxiv.org/abs/1603.06273}{[arXiv:1603.06273]}.

\bibitem{amp1}S. Giombi, C. Sleight, M. Taronna, \emph{JHEP} \textbf{06}
(2018) 030 \href{https://arxiv.org/abs/1708.08404}{[arXiv:1708.08404]}.

\bibitem{amp2}C. Sleight, M. Taronna, \emph{JHEP} \textbf{01} (2018)
060 \href{https://arxiv.org/abs/1708.08668}{[arXiv:1708.08668]}.

\bibitem{amp3}D. Ponomarev, E. Sezgin, E. Skvortsov, \emph{JHEP}
\textbf{11} (2019) 138 \href{https://arxiv.org/abs/1904.01042}{[arXiv:1904.01042]}.

\bibitem{amp4}B. Nagaraj, D. Ponomarev, \emph{Phys.Rev.Lett.} \textbf{122}
(2019) 10, 101602 \href{https://arxiv.org/abs/1811.08438}{[arXiv:1811.08438]}.

\bibitem{amp5}R. de Mello Koch, A. Jevicki, K. Suzuki, J. Yoon, \emph{JHEP}
\textbf{03} (2019) 133, \href{https://arxiv.org/abs/1810.02332}{[arXiv:1810.02332]}.

\bibitem{amp6}B. Nagaraj, D. Ponomarev, \emph{JHEP} \textbf{06} (2020)
068 \href{https://arxiv.org/abs/1912.07494}{[arXiv:1912.07494]}.

\bibitem{amp7}B. Nagaraj, D. Ponomarev, \emph{JHEP} \textbf{08} (2020)
08, 012 \href{https://arxiv.org/abs/2004.07989}{[arXiv:2004.07989]}.

\bibitem{chir1}D. Ponomarev, E.D. Skvortsov, \emph{J.Phys.A} \textbf{50}
(2017) 9, 095401 \href{https://arxiv.org/abs/1609.04655}{[arXiv:1609.04655]}.

\bibitem{chir2}D. Ponomarev, \emph{JHEP} \textbf{12} (2016) 117 \href{https://arxiv.org/abs/1611.00361}{[arXiv:1611.00361]}.

\bibitem{chir3}D. Ponomarev, \emph{JHEP} \textbf{12} (2017) 141 \href{https://arxiv.org/abs/1710.00270}{[arXiv:1710.00270]}.

\bibitem{chir4}E.D. Skvortsov, T. Tran, M. Tsulaia, \emph{Phys.Rev.Lett.}
\textbf{121} (2018) 3, 031601 \href{https://arxiv.org/abs/1805.00048}{[arXiv:1805.00048]}.

\bibitem{chir5}E. Skvortsov, T. Tran, \emph{JHEP} \textbf{07} (2020)
021 \href{https://arxiv.org/abs/2004.10797}{[arXiv:2004.10797]}.

\bibitem{chir6}E. Skvortsov, T. Tran, M. Tsulaia, \emph{Phys.Rev.D}
\textbf{101} (2020) 10, 106001 \href{https://arxiv.org/abs/2002.08487}{[arXiv:2002.08487]}.

\bibitem{misuna}N.G. Misuna, \emph{Phys.Lett.B} \textbf{798} (2019)
134956 \href{https://arxiv.org/abs/1905.06925}{[arXiv:1905.06925]}.

\bibitem{mass1}Yu.M. Zinoviev, \emph{Nucl.Phys.B} \textbf{808} (2009)
185-204 \href{https://arxiv.org/abs/0808.1778}{[arXiv:0808.1778]}.

\bibitem{key-21}D.S. Ponomarev, M.A. Vasiliev, \emph{Nucl.Phys.B}
\textbf{839} (2010) 466-498 \href{https://arxiv.org/abs/1001.0062}{[arXiv:1001.0062]}.

\bibitem{key-23}I.L. Buchbinder, T.V. Snegirev, Yu.M. Zinoviev, \emph{Phys.Lett.B}
\textbf{716} (2012) \href{https://arxiv.org/abs/1207.1215}{[arXiv:1207.1215]}.

\bibitem{key-24}Yu.M. Zinoviev, \emph{J.Phys.A} \textbf{49} (2016)
9, 095401 \href{https://arxiv.org/abs/1509.00968}{[arXiv:1509.00968]}.

\bibitem{key-25}I.L. Buchbinder, T.V. Snegirev, Yu.M. Zinoviev, \emph{JHEP}
\textbf{08} (2016) \href{https://arxiv.org/abs/1606.02475}{[arXiv:1606.02475]}.

\bibitem{key-26}I.L. Buchbinder, M.V. Khabarov, T.V. Snegirev, Yu.M.
Zinoviev, \emph{Nucl.Phys.B} \textbf{942} (2019) 1-29 \href{https://arxiv.org/abs/1901.09637}{[arXiv:1901.09637]}.

\bibitem{key-27}M.V. Khabarov, Yu.M. Zinoviev, \emph{Nucl.Phys.B}
\textbf{948} (2019) 114773 \href{https://arxiv.org/abs/1906.03438}{[arXiv:1906.03438]}.

\bibitem{mass2}M.V. Khabarov, Yu.M. Zinoviev, \emph{Nucl.Phys.B}
\textbf{953} (2020) 114959 \href{https://arxiv.org/abs/2001.07903}{[arXiv:2001.07903]}.

\bibitem{unf1}M.A. Vasiliev, \emph{Annals Phys.} \textbf{190} (1989)
59-106.

\bibitem{Lyakh1}S.L. Lyakhovich, A.A. Sharapov, \emph{JHEP} \textbf{0602}
(2006) 007 \href{https://arxiv.org/abs/hep-th/0512119}{[hep-th/0512119]}.

\bibitem{Lyakh2}D.S. Kaparulin, S.L. Lyakhovich, A.A. Sharapov, \emph{Int.J.Mod.Phys.A}
\textbf{26} (2011) 1347-1362 \href{https://arxiv.org/abs/1012.2567}{[arXiv:1012.2567]}.

\bibitem{Lyakh3}D.S. Kaparulin, S.L. Lyakhovich, A.A. Sharapov, \emph{SIGMA}
\textbf{8} (2012) 021 \href{https://arxiv.org/abs/1112.1860}{[arXiv:1112.1860]}.

\bibitem{sigma_cohomol}O.V. Shaynkman, M.A. Vasiliev, \emph{Theor.Math.Phys.}
\textbf{123} (2000) 683\textendash 700 \href{https://arxiv.org/abs/hep-th/0003123}{[hep-th/0003123]}.

\bibitem{Fronsdal}C. Fronsdal, \emph{Phys.Rev.D} \textbf{20} (1979)
848-856.

\bibitem{AdS_propagator}C.P. Burgess, C.A. Lütken, \emph{Phys.Lett.B}
\textbf{153} (1985) 3, 137-141.
\end{thebibliography}
\end{document}